\documentclass[11pt]{article}

\usepackage{amsmath,amssymb}
\usepackage{slashed}
\usepackage{xcolor} %Define colors
\usepackage{graphicx}
\usepackage{mathtools}

\usepackage{amstext}
\usepackage{amsmath}
\usepackage{multirow}
\usepackage{hhline}
\usepackage{amssymb}
\usepackage{hyperref}
\usepackage[all]{hypcap}
\usepackage{xspace}
\usepackage{endnotes}
\usepackage{rotating} % for long tables
\usepackage{subfigure}
\usepackage{epstopdf}
\usepackage{graphicx}
\usepackage{color}
\usepackage[export]{adjustbox}
\usepackage{progressbar}
\usepackage{caption}
\usepackage{braket}

\usepackage{url}
\usepackage{cite}

\newcommand{\beq}{\begin{equation}}
\newcommand{\eeq}{\end{equation}}
\newcommand{\bea}{\begin{eqnarray}}
\newcommand{\eea}{\end{eqnarray}}

\newcommand{\fref}[1]{Fig.~\ref{fig:#1}} 
\newcommand{\eref}[1]{Eq.\eqref{eqn:#1}} 

\newcommand{\sref}[1]{Sec.~\ref{sec:#1}}
\newcommand{\ssref}[1]{Sec.~\ref{subsec:#1}}

\graphicspath{{./Figures/}}
\input{definitions.sty}

\begin{document}

\vspace*{-2cm}
\begin{flushright}
FERMILAB-PUB-14-396-T, LPT-Orsay-14-76, nuhep-th/14-06
\vspace*{2mm}
\end{flushright}

\begin{center}
\vspace*{5mm}

\vspace{.01cm}
{\Large 
{\bf Technical Note for 8D Likelihood Effective Higgs\\ Couplings Extraction Framework
in $h\to 4\ell$} \\
\vspace{.15cm}
{\it Part I: From Generator Level to Detector Level}
} \\
\vspace{.5cm}
{\bf Yi Chen$\,^{a}$,~Emanuele Di Marco$\,^{a}$,~Joe Lykken$\,^{b}$,\\
~Maria Spiropulu$\,^{a}$,~Roberto Vega-Morales$\,^{b,c,d}$,~Si Xie$\,^{a}$}

\vspace*{.3cm} 
$^a$Lauritsen Laboratory for High Energy Physics,\\ California Institute of Technology, Pasadena, CA, 92115, USA\\
$^b$Theoretical Physics Department, Fermilab, P.O.~Box 500, Batavia, IL 60510, USA\\
$^c$Laboratoire de Physique Th\'{e}orique, CNRS - UMR 8627, Universit\'{e} Paris-Sud, Orsay, France\\
$^d$Department of Physics and Astronomy, Northwestern University, Evanston, IL 60208, USA 
\vspace*{.2cm} 

\end{center}

\vspace*{.5mm}
\begin{abstract}\noindent\normalsize
In this technical note we present technical details on various aspects of the framework introduced in~\cite{Chen:2014pia} aimed at extracting effective Higgs couplings in the $h\to 4\ell$ `golden channel'.~Since it is the primary feature of the framework, we focus in particular on the convolution integral which takes us from `truth' level to `detector' level and the numerical and analytic techniques used to obtain it.~We also briefly discuss other aspects of the framework.
\end{abstract}

%%%%%%%%%%%%%%%%%%%%%%%%%%%%%%%%%%
\section{Introduction} \label{sec:intro} 
It is well known that the $h\to 4\ell$ ($4\ell = 2e2\mu, 4e, 4\mu$) `golden channel' is a powerful means of studying the Higgs couplings to neutral electroweak gauge bosons and various methods have long been proposed for studying it~\cite{Nelson:1986ki,Soni:1993jc,Chang:1993jy,Barger:1993wt,Arens:1994wd} and more recently~\cite{Choi:2002jk,Buszello:2002uu,Godbole:2007cn,Kovalchuk:2008zz,Cao:2009ah,Gao:2010qx,DeRujula:2010ys,Gainer:2011xz,Coleppa:2012eh,Bolognesi:2012mm,Stolarski:2012ps,Chen:2012jy,Boughezal:2012tz,Belyaev:2012qa,Avery:2012um,Campbell:2012ct,Campbell:2012cz,Chatrchyan:2012sn,Chatrchyan:2012ufa,Chatrchyan:2012jja,Modak:2013sb,Sun:2013yra,Gainer:2013rxa,Artoisenet:2013puc,Anderson:2013fba,Chen:2013waa,Buchalla:2013mpa,Chen:2013ejz,Gainer:2014hha,Chen:2014gka}.~Though `truth' level (or generator) studies of the golden channel give a good approximate estimate of the expected sensitivity to the Higgs $ZZ$, $Z\gamma$, and $\gamma\gamma$ couplings~\cite{Chen:2014gka}, when analyzing data obtained at the LHC (or future colliders) a detector level likelihood which accounts for the various detector effects is necessary.~Since generally detector level likelihoods are obtained via the use of Monte Carlo methods, it becomes difficult to obtain the full multi-dimensional likelihood for the $4\ell$ final state.~Typically one needs to fill large multi-dimensional templates that require an impractical amount of computing time.~There are also potential collateral binning and `smoothing' side-effects often associated with these methods.~In the case of the golden channel this necessitates the use of kinematic discriminants which `collapse' the fully multi-dimensional likelihood into two or perhaps three detector level observables~\cite{Anderson:2013fba}.~This approach is normally taken to facilitate the inclusion of detector effects, but is not optimal when fitting to a large number of parameters simultaneously~\cite{Bolognesi:2012mm,gaotalk}.~This is unfortunate in the case of the golden channel where in principle there are twelve observables which can be used to extract a large number of parameters at once, including their correlations.~It would be satisfying and useful to have a framework which is free of these issues and capable of utilizing all available information in the four lepton final state at detector level.

This is accomplished in our framework~\cite{Chen:2014pia} by performing an explicit convolution of the generator (`truth') level probability density, formed out of analytic expressions for the signal and background differential cross sections, with a transfer function which encapsulates the relevant detector effects.~This can be represented schematically as follows,
\begin{eqnarray}
\label{eqn:conv}
&& P(\vec{X}^\mathrm{R} | \vec{\mathcal{A}} ) = \int P(\vec{X}^G | \vec{\mathcal{A}} ) T({\vec{X}^{R} | \vec{X}^{G}}) d\vec{X}^G.
\end{eqnarray}
Here we take $\vec{X}$ to represent the full set of center of mass variables, of which there are twelve in the golden channel, to be discussed more below, and $\vec{\mathcal{A}}$ represents some set of lagrangian parameters~\cite{Chen:2014pia}.~The transfer function~$T({\vec{X}^{R} | \vec{X}^{G}})$ takes us from generator (G) level to reconstructed (R) (or detector) level observables and represents the probability of reconstructing the observables $\vec{X}^R$ given the generator level observable $\vec{X}^G$.~It is treated as a function of $\vec{X}^R$ which takes $\vec{X}^G$ as input.~As will be described more below, once the integration in~\eref{conv} is performed we must then normalize over all twelve reconstructed level observables to obtain the detector level probability density function (\emph{pdf}).~After performing this 12-dimensional integration and normalizing, we are left with a \emph{pdf} from which \emph{we construct an un-binned twelve-dimensional detector level likelihood} which is a continuous function of the effective couplings (or Lagrangian parameters) and takes as its input, up to twelve reconstructed, detector-level center of mass observables.~In the current implementation~\cite{Chen:2014pia} we average over the four production variables to reduce the systematic uncertainties, thus obtaining an eight-dimensional likelihood in terms of just decay observables.~However, this step is in principle not necessary.

We have performed the integration in~\eref{conv} for both the $h\to4\ell$ signal as well as the dominant $q\bar{q}\to4\ell$ background (computed in~\cite{Chen:2012jy,Chen:2013ejz,Vega-Morales:2013dda}) and emphasize that it has not been done via Monte Carlo methods, but instead by a combination of numerical and analytical techniques to be discussed in detail below.~With these detector level \emph{pdfs} in hand we can go on to perform fast and accurate multi-parameter extractions on data obtained at colliders in the $4\ell$ channel as demonstrated in~\cite{Chen:2014pia} as well as in recent CMS studies~\cite{CMS-PAS-HIG-14-014}.~In this technical note we present details on various aspects of how the convolution integral is performed and briefly discuss other aspects of the framework.~Further details can be found in accompanying studies~\cite{Chen:2012jy,Chen:2013ejz,Vega-Morales:2013dda,Chen:2014gka,Chen:2014pia,Thesis} as well as~\cite{CMS-PAS-HIG-14-014}.

%%%%%%%%%%%%%%%%%%%%%%%%%%%%%%%%
%%%%%%%%%%%%%%%%%%%%%%%%%%%%%%%%
%%%%%%%%%%%%%%%%%%%%%%%%%%%%%%%%
\section{From `Truth' Level to `Detector' Level}
\label{Section:MultiDimensionalOverview-Convolution}

We now describe in more detail how the convolution integral in~\eref{conv} is performed.~We first review the twelve center of mass variables present in the $4\ell$ system before describing the various numerical and analytic techniques used in the integration which takes us from `truth' level to detector level.

%%%%%%%%%%%%%%%%%%%%%%%%%
\subsection{Center of Mass Observables}
\label{subsec:events}

The twelve variables are discussed in more detail in~\cite{Chen:2012jy,Stolarski:2012ps,Chen:2013ejz,Chen:2014gka,Chen:2014pia}, but are listed here for convenience as they will be used extensively in what follows.~The invariant masses are defined as:
\begin{itemize}
\item $\sqrt{\hat{s}} \equiv M_{4\ell} \equiv m_h$ -- The invariant mass of the four lepton system or the Higgs mass in case of signal. 
\item $M_{1}$ -- The invariant mass of the lepton pair system which reconstructs closest to the $Z$ mass. 
\item $M_{2}$ -- The invariant mass of the other lepton pair system and interpreted as $M_2 < M_1$.~This condition holds as long as $\sqrt{\hat{s}} \lesssim 2 M_Z$. 
\end{itemize}
These invariant masses are all independent subject to the constraint $(M_1 + M_2) \leq \sqrt{\hat{s}}$ and serve as the most strongly discriminating observables between different signal hypothesis as well as between signal and background.~Note also that the $4e/4\mu$ final state can be reconstructed in two different ways due to the identical final state interference.~This is a quantum mechanical effect that occurs at the amplitude level and thus both reconstructions are valid.~The definitions $M_1$ and $M_2$ remain unchanged however.~The angular variables are defined as:
\begin{itemize}
\item $\Theta$ -- The production angle between the momentum vectors of the lepton pair which reconstructs to $M_1$ and the total $4\ell$ system momentum.
\item $\theta_{1,2}$ -- Polar angle of the momentum vectors of $e^-,\mu^-$ in the lepton pair rest frame.
\item $\Phi_1$ -- The angle between the plane formed by the $M_1$ lepton pair and the `production plane' formed out of the momenta of the incoming partons and the momenta of the two lepton pair systems.
\item $\Phi$ -- The angle between the decay planes of the final state lepton pairs in the rest frame of the $4\ell$ system.
\end{itemize}
We group the angular variables as follows $\vec{\Omega} = (\Theta, \cos\theta_1, \cos\theta_2, \Phi_1, \Phi)$.~These angular variables are useful in aiding to distinguish different signal hypothesis and in particular between those with different CP properties, as well as in discriminating signal from background.~Lastly, we have production variables associated with the initial partonic state four momentum:
\begin{itemize}
\item $\vec{p}_T \equiv (p_T \cos\phi_{4\ell}, p_T \sin\phi_{4\ell})$ -- The momentum in the transverse direction of the $4\ell$ system.
\item $Y$ -- Defined as the motion of the $4\ell$ system along the longitudinal direction.
\item $\phi$ -- Defines a global rotation of the event in the $4\ell$ rest frame.
\end{itemize}
%

%%%%%%%%%%%%%%%%%%%%%%%
\subsection{Changing Variables for Background \emph{pdf}}
\label{subsec:bg_change_var}

Beginning from~\eref{conv} we first discuss the construction of the background detector level \emph{pdf}.~The construction of the signal will be discussed separately as there is a subtle, but important, difference in performing the convolution.~Since there are no undetermined parameters in the background the generator and detector-level (un-normalized) differential cross sections are given simply by $P_B(\vec{X}^G)$ and $P_B(\vec{X}^R)$ respectively and the convolution can be written as,
\begin{eqnarray}
\label{eqn:reco_bg}
P_B(\vec{X}^\mathrm{R}) &=& \int P_B(\vec{X}^G ) T({\vec{X}^{R} | \vec{X}^{G}}) d\vec{X}^G .
\end{eqnarray}
The set of variables $\vec{X} \equiv (\vec{p}_{T},Y,\phi,\hat{s},M_{1},M_{2},\vec{\Omega})$ exhausts the twelve degrees of freedom (note that $\vec{p}_T$ has 2 components and $\vec{\Omega}$ contains 5 angles) available to the four (massless) final state leptons.~The differential volume element is given by $d\vec{X} = d\hat{s} dM^2_{1}dM^2_{2}d\vec{\Omega} \cdot d\vec{p}_{T}dYd\phi$.

To perform this convolution with the transfer function we must first transform to the basis in which the detector smearing of the lepton momenta is parameterized.~This requires that we first transform from the basis of the twelve center of mass variables defined in~\ssref{events} to the three momentum basis for the four final state leptons.~This can be represented as follows,
\begin{eqnarray}
\label{eqn:reco_pdf2A}
P_B(\vec{X}^\mathrm{R}) &=& \int P_B(\vec{X}^G ) T({\vec{X}^{R} | \vec{X}^{G}}) d\vec{X}^G \nonumber \\
&=& \int P_B(\vec{X}^G )T({\vec{P}^{R} | \vec{P}^{G}}) \frac{|{\bf J}^{\vec{P}}_G|}{|{\bf J}^{\vec{P}}_R|} d\vec{P}^G,
\end{eqnarray}
where the differential volume element is now given by,
\begin{eqnarray}
d\vec{P}^G = \prod \limits_{i=1}^4 d\vec{p}^{~G}_i,
\end{eqnarray}
and $\vec{p}^{~G}_i$ is the generator level three momentum of the i'th lepton.~The $|{\bf J}^{\vec{P}}_G|$ is the Jacobian associated with the twelve dimensional change of variables from $\vec{X}^G \rightarrow \vec{P}^G$ in the differential volume element.~The $|{\bf J}^{\vec{P}}_R|$ arises from the change of variables $\vec{X}^R \rightarrow \vec{P}^R$ in the transfer function (remembering $T({\vec{X}^{R} | \vec{X}^{G}})$ is treated as a function of $\vec{X}^R$) which we loosely also refer to as a Jacobian, as we will do for all subsequent change of variables to follow.~Ideally to find these Jacobian factors one should construct the $12\times12$ matrix associated with these transformations and then calculate the determinant.~However, since these transformations are highly non-linear and must be performed for each point in phase space, this is untenable analytically.~We therefore implement a numerical algorithm to calculate these factors for each phase space point which will be discussed in more detail in~\ssref{jacobians}.

Since we make the assumption that detector smearing will only affect the component of the lepton momentum parallel to the direction (${p_i}_{||}$) of motion and not the two components perpendicular to the direction of motion ($\vec{p_i}_{\perp}$) (which are zero at generator level) we find it convenient to decompose the lepton three momenta $\vec{p}_i$ in terms of ${p_i}_{||}$ and $\vec{p_i}_{\perp}$.~Note that this assumption is equivalent to assuming angular resolution effects due to detector smearing can be neglected,~which is an excellent approximation for the LHC detectors~\cite{CMSperformance2,CMS-DP-2013-003}.~In the (${p_i}_{||}, \vec{p_i}_{\perp}$) basis only the transfer function associated with ${p_i}_{||}$ is non-trivial while the one associated with the perpendicular components can be represented simply as a delta function for each perpendicular direction, thus allowing for trivial integration over the eight $\vec{p_i}_{\perp}$ variables. 

The differential volume element can now be written as,
\begin{eqnarray}
d\vec{P}^{G} = \prod\limits_{i=1}^4 d\vec{p}^{~G}_i = 
\prod\limits_{i=1}^4 d\vec{p_i}^G_{\perp} d{p_i}^G_{||} . 
\end{eqnarray}
We then use the property of the transfer function that it is explicitly parametrized in terms of the ratio of reconstructed and generator level momentum components along the direction of motion to again change variables as follows,
\begin{eqnarray}
\label{eqn:reco_pdf2}
P_B(\vec{X}^\mathrm{R} ) 
&=& \int P_B(\vec{X}^G )T({\vec{P}^{R} | \vec{P}^{G}}) \frac{|{\bf J}^{\vec{P}}_G|}{|{\bf J}^{\vec{P}}_R|} \prod\limits_{i=1}^4 d\vec{p_i}^G_{\perp} d{p_i}^G_{||}\nonumber \\
&=& \int P_B(\vec{X}^G )T(\vec{c}~| \vec{P}^{G}) \frac{|{\bf J}^{\vec{P}}_G|}{|{\bf J}^{\vec{P}}_R|} \frac{|{\bf J}^{\vec{c}}_G|}{|{\bf J}^{\vec{c}}_R|} \prod\limits_{i=1}^4 dc_i d\vec{p_i}_{\perp}^G,
\end{eqnarray} 
where we have defined $c_i = {p_i}^R_{||}/{p_i}^G_{||}$ and $\vec{c} = (c_1, c_2, c_3, c_4)$.~The components of the Jacobians $|{\bf J}^{\vec{c}}_R|$ and $|{\bf J}^{\vec{c}}_G|$ which take us from ${p_i}^R_{||} \rightarrow c_i$ and ${p_i}^G_{||} \rightarrow c_i$ variables are obtained trivially as $|1/{p_i}^G_{||}|$ and $|c_i/{p_i}^G_{||}|$ for the transfer function (which is now a function of $\vec{c}$) and the differential volume element respectively.~Finally, we use the fact that $c_{1} c_{2} = (M_{1}^R/M_{1}^G)^2$ and $c_{3} c_{4} = (M_{2}^R/M_{2}^G)^2$ to eliminate $c_2$ and $c_4$ and make a final change of variables to the basis in which we perform the explicit four dimensional integration,
\bea
\label{eqn:reco_pdf3}
P_B(\vec{X}^\mathrm{R} ) 
&=& \int P_B(\vec{X}^G )T(\vec{c}~| \vec{P}^{G}) 
\frac{|{\bf J}^{\vec{P}}_G|}{|{\bf J}^{\vec{P}}_R|} 
\frac{|{\bf J}^{\vec{c}}_G|}{|{\bf J}^{\vec{c}}_R|} 
\prod\limits_{i=1}^4 dc_i \\
&=& \int P_B(\vec{X}^G )T(\vec{c}~| \vec{P}^{G}) 
\times 
\frac{|{\bf J}^{\vec{P}}_G|}{|{\bf J}^{\vec{P}}_R|} 
\frac{|{\bf J}^{\vec{c}}_G|}{|{\bf J}^{\vec{c}}_R|} 
|{\bf J}^{\vec{M}}_B| 
\cdot
dc_1 dc_3 d{M_1^2}^G d{M_2^2}^G, \nonumber 
\eea
where in the first line in~\eref{reco_pdf3} we have implicitly used the delta functions in the transfer function to perform the eight dimensional integration over $\vec{p_i}_{\perp}^G$.~The Jacobian $|{\bf J}^{\vec{M}}_B|$ is obtained analytically from the change of variables $c_2, c_4 \rightarrow {M_1^G}^2 {M_2^G}^2$ by observing that,
\bea
(p_1^G + p_2^G)^2 &=& {M_1^G}^2,  \nonumber\\
(p_3^G + p_4^G)^2 &=& {M_2^G}^2, \nonumber\\
(p_1^R + p_2^R)^2 &=& (c_1 p_1^G + c_2 p_2^G)^2 = {M_1^R}^2, \nonumber\\
(p_3^R + p_4^R)^2 &=& (c_3 p_3^G + c_4 p_4^G)^2 = {M_2^R}^2 .
\eea
Assuming that the leptons are massless (an excellent approximation for muons and electrons) and expanding out the equations we arrive at,
\bea
{M_1^R}^2 = 2 c_1 c_2 p_1^G p_2^G = c_1 c_2 {M_1^G}^2, \nonumber \\
{M_2^R}^2 = 2 c_3 c_4 p_3^G p_4^G = c_3 c_4 {M_2^G}^2,
\eea
From here we solve for the smearing factors $c_2$ and $c_4$, 
\bea
c_2 &=& \dfrac{1}{c_1}\dfrac{{M_1^R}^2}{{M_1^G}^2} \equiv \dfrac{1}{c_1} R_{12}\nonumber\\
c_4 &=& \dfrac{1}{c_3}\dfrac{{M_2^R}^2}{{M_2^G}^2} \equiv \dfrac{1}{c_3} R_{34}.
\label{eqn:c2c4Expressions}
\eea
from which the Jacobian elements are easily computed as,
\bea
d c_2 = -\dfrac{1}{c_1} \dfrac{{M_1^R}^2}{{M_1^G}^4} d {M_1^G}^2\nonumber\\
d c_4 = -\dfrac{1}{c_3} \dfrac{{M_2^R}^2}{{M_2^G}^4} d {M_2^G}^2.
\eea
This gives finally for $|{\bf J}_B^{\vec{M}}|$,
\bea
|{\bf J}_B^{\vec{M}}| = \dfrac{1}{c_1} \dfrac{{M_1^R}^2}{{M_1^G}^4} \dfrac{1}{c_3} \dfrac{{M_2^R}^2}{{M_2^G}^4}.
\eea
We thus see in~\eref{reco_pdf3} that what started out as a twelve dimensional integral has been reduced to a much more manageable integration over four variables.~The details and validation of this four dimensional integration will be presented in~\sref{MultiDimensionalDetailsIntegration}, but first we discuss the change of variables in signal case.

%%%%%%%%%%%%%%%%%%%%%%%%%%%%%%%
\subsection{Changing Variables for Signal \emph{pdf}}
\label{subsec:sig_change_var}
To construct the detector level signal \emph{pdf}, which is now a function of the effective couplings $\vec{\mathcal{A}}$, we follow the same procedure as for the background through the second line in~\eref{reco_pdf2} to obtain,
\begin{eqnarray}
\label{eqn:reco_pdf_sig}
P_S(\vec{X}^\mathrm{R} | \vec{\mathcal{A}}) &=& 
\int P_S(\vec{X}^G | \vec{\mathcal{A}}) T(\vec{c}~| \vec{P}^{G}) 
\times \frac{|{\bf J}^{\vec{P}}_G|}{|{\bf J}^{\vec{P}}_R|} 
\frac{|{\bf J}^{\vec{c}}_G|}{|{\bf J}^{\vec{c}}_R|} 
\prod\limits_{i=1}^4 dc_i d\vec{p_i}_{\perp}^G. 
\end{eqnarray}
In contrast to the background however, we now perform the following change of variables, 
\begin{eqnarray}
\label{eqn:reco_pdf_sig2}
P_S(\vec{X}^\mathrm{R} | \vec{\mathcal{A}}) = 
\int P_S(\vec{X}^G | \vec{\mathcal{A}}) T(\vec{c}~| \vec{P}^{G}) \nonumber 
\times 
\frac{|{\bf J}^{\vec{P}}_G|}{|{\bf J}^{\vec{P}}_R|} 
\frac{|{\bf J}^{\vec{c}}_G|}{|{\bf J}^{\vec{c}}_R|} 
|{\bf J}^{\vec{M}}_S| 
\cdot
d\hat{s}^G dc_1 d{M_1^2}^G d{M_2^2}^G,
\end{eqnarray}
where again we have implicitly used the delta functions in the transfer function to perform the eight dimensional integration over $\vec{p_i}_{\perp}^G$.~Here $|{\bf J}^{\vec{M}}_S|$ is the Jacobian obtained analytically in the change of variables $c_2, c_3, c_4 \rightarrow \hat{s}^G, {M_1^2}^G, {M_2^2}^G$ by using~\eref{c2c4Expressions} and the following relation for $\hat{s}^G$,
\bea
\hat{s}^G = \sum_{i > j} c_i^{-1} c_j^{-1} {M_{ij}^R}^2.
\eea
This allows us to write down the transformation matrix as,
\bea
\hat{M} =
\begin{bmatrix}
\dfrac{\partial {M_1^G}^2}{\partial c_2} & \dfrac{\partial \hat{s}^G}{\partial c_2} & 0 \\
0 & \dfrac{\partial \hat{s}^G}{\partial c_3} & \dfrac{\partial {M_2^G}^2}{\partial c_3} \\
0 & \dfrac{\partial \hat{s}^G}{\partial c_4} & \dfrac{\partial {M_2^G}^2}{\partial c_4}
\end{bmatrix} ,
\eea
from which the Jacobian can be obtained by,
\bea
|{\bf J}_S^{\vec{M}}| = \dfrac{1}{|det(\hat{M})|}.
\eea
As discussed in more detail in~\cite{Chen:2014pia}, the `truth' level $\hat{s}^G$ spectrum for the signal is given by a delta function\footnote{Note that the delta function approximation is taken for the recently discovered $\sim 125$~GeV Higgs boson, but need not be imposed if a new heavy scalar with a large width is discovered in the future.~In this case, the $\hat{s}$ spectrum can be treated similarly to the background case.} $\propto \delta(\hat{s}^G-m_h^{2})$ (where $m_h$ is the $\emph{generated}$ Higgs mass), enabling us to perform the integration over $d\hat{s}^G$.~Thus, we have for the final signal detector level \emph{pdf},
\bea
\label{eqn:reco_pdf_sig3}
P_S(\vec{X}^\mathrm{R} | \vec{\mathcal{A}}) = 
\int P_S(\vec{X}^G | \vec{\mathcal{A}}) T(\vec{c}~| \vec{P}^{G}) 
\times 
\frac{|{\bf J}^{\vec{P}}_G|}{|{\bf J}^{\vec{P}}_R|} 
\frac{|{\bf J}^{\vec{c}}_G|}{|{\bf J}^{\vec{c}}_R|} 
|{\bf J}^{\vec{M}}_S| 
\cdot
dc_1 d{M_1^2}^G d{M_2^2}^G\Big|_{\hat{s}^G=m_h^2} .
\eea
We note that the delta function in $\hat{s}^G$ introduced additional complications which are computationally non-trivial when including detector resolution effects.~This is because the delta function in $\hat{s}^G$ places an additional constraint when performing the ${M_1^2}^G$, ${M_2^2}^G$ integration which must be properly taken into account.~We discuss these issues in more detail in~\ssref{MultiDimensionalDetailsSignal}, but first we show how the $12\times12$ jacobians ${\bf J}_G^{\vec{P}}$ and ${\bf J}_R^{\vec{P}}$ are computed numerically.

%%%%%%%%%%%%%%%%%%%%%%%%%%
\subsection{Calculation of the ${\bf J}_G^{\vec{P}}$ and ${\bf J}_R^{\vec{P}}$ Jacobian Factors}
\label{subsec:jacobians}

We now turn to the $12\times 12$ Jacobians ${\bf J}_G^{\vec{P}}$ and ${\bf J}_R^{\vec{P}}$ which take us from the center of mass basis to the `lepton-smearing basis'.~More explicitly these Jacobians define the twelve-dimensional (for known lepton masses) transformation,
\bea
\vec{X} \equiv (\vec{p}_{T},Y,\phi,\hat{s},M_{1},M_{2},\vec{\Omega}) 
\Longrightarrow \vec{P} \equiv ({p_1}_{||}, {p_2}_{||}, {p_3}_{||}, {p_4}_{||}, \vec{p_1}_{\perp}, \vec{p_2}_{\perp}, \vec{p_3}_{\perp}, \vec{p_4}_{\perp}).
\eea
Ideally one should simply work out the $12\times12$ matrix and calculate the discriminant.~However, since it involves many boosts and trigonometric functions to perform this non-linear transformation, it is not possible to obtain analytically.~Furthermore, since the components of $\vec{P}$ in principle depend on the particular point in $\vec{X}$ this transformation must be obtained for each point in phase space.~Therefore we take another approach and calculate the factor numerically.

The Jacobian factors have a simple geometrical interpretation; they can be interpreted as the ratio of infinitesimal volume elements before and after the change of basis.~This is illustrated schematically in~\fref{JacobiansGeometricalInterpretation} where $V$ and $V^\prime$ represent the infinitesimal volume elements around a given point in the two different coordinate systems $\vec{X}$ and $\vec{P}$ respectively.~Note that even though the lepton momentum basis is dependent on the particular point in $\vec{X}$, this does not affect the final calculation of the Jacobian since:
\begin{itemize}
\item As we scan through the different $\vec{X}$ and transform them to the basis $\vec{P}$, the parallel component directions in $\vec{P}$ line up during the integration.~Thus, we can use the same basis for each point in $\vec{X}$ during the integration over ${p_i}_{||}$.
\item The freedom of choice in the perpendicular components is irrelevant since the delta function in $\vec{p_i}_{\perp}$ constrains these directions to be fixed.~Without this constraint, one would need to carefully line up the perpendicular directions during the convolution integral.
\item The lepton vector basis $\vec{P}$ between different points in the $\vec{X}$ basis are related by simple
rotations which leave the volume, and thus the Jacobian, invariant.
\end{itemize}
These are the key features which make the integration possible and allow us to, point by point in the phase space, numerically build an infinitesimal 12-dimensional cube with volume $V$ in the basis $\vec{X}$ and then transform it to the basis $\vec{P}$ where the volume $V^\prime$ of the resulting hyper-parallelepiped is calculated.~Calculating the volume of this 12-dimensional parallelepiped can be done using various readily available algorithms~\cite{Bradie:943117} allowing us to obtain the volume $V^\prime$.~We implement a simple algorithm where the hyper-parallelepiped is transformed into a hyper-cube with equal volume $V^\prime$ as follows:
\begin{itemize}
\item Choose any vector and pair it with a second vector.
\item Subtract out the parallel component of the first vector from the second vector.
\item Take a third vector, subtract out the parallel components of the first vector,\\ as well as that of the modified second vector.
\item Repeat the process for all remaining sides of the parallelepiped until a hyper-`cube'\\ can be constructed.
\item Product of length of all the edges now gives the volume $V^\prime$.
\end{itemize}
This is equivalent to calculating the determinant of the $12\times12$ transformation matrix, but
conceptually easier to visualize.

This calculation of this Jacobian factor can be validated with toy distributions.~For any given test distribution $f(\vec{X})$, we can generate events in two different bases and compare the distribution with one weighted by the Jacobian factor.~To see this consider,
\bea\label{eqn:Jcheck}
f(\vec{X}) d\vec{X} = f(\vec{X}) |{\bf J^{\vec{P}}}| d\vec{P},
\eea
where ${\bf J^{\vec{P}}}$ represents either ${\bf J}_G^{\vec{P}}$ or ${\bf J}_R^{\vec{P}}$ .~Starting from~\eref{Jcheck} one can generate events from the left hand side and right hand side separately and then compare the two datasets.~They should be identical if the Jacobian has been calculated correctly.~Since the Jacobians arise from the change of variable and do not depend on other details of the integration, one has the freedom to choose different toy integrands to validate the calculation of this Jacobian factor.~The toy function used is as follows:
\begin{eqnarray}
f(\vec{X}) =
\begin{dcases}
1, & \text{if~~} 100 \text{~GeV} < \sqrt{\hat{s}} < 140 \text{~GeV}, \\
 & 4 \text{~GeV} < M_{1,2} < 100 \text{~GeV},~~|\vec{p}_i| < 100 \text{~GeV} \\
 & |Y| < 4,~~|\vec{p}_T| < 100 \text{~GeV} \\
0, & \text{otherwise}
\end{dcases}
\end{eqnarray}
The result of the validation is shown in~\fref{MultiDimensionalDetails-Jacobians-12DValidation} where we have decomposed the four lepton system transverse momentum into its components as $\vec{p}_{T} = (p_T \cos\phi_{4\ell}, p_T \sin\phi_{4\ell})$.~We see excellent agreement between events generated in the two different basis for all twelve variables, indicating that the calculation is performed correctly.
\begin{figure}
\centering
\includegraphics[width=0.45\textwidth, angle=270]{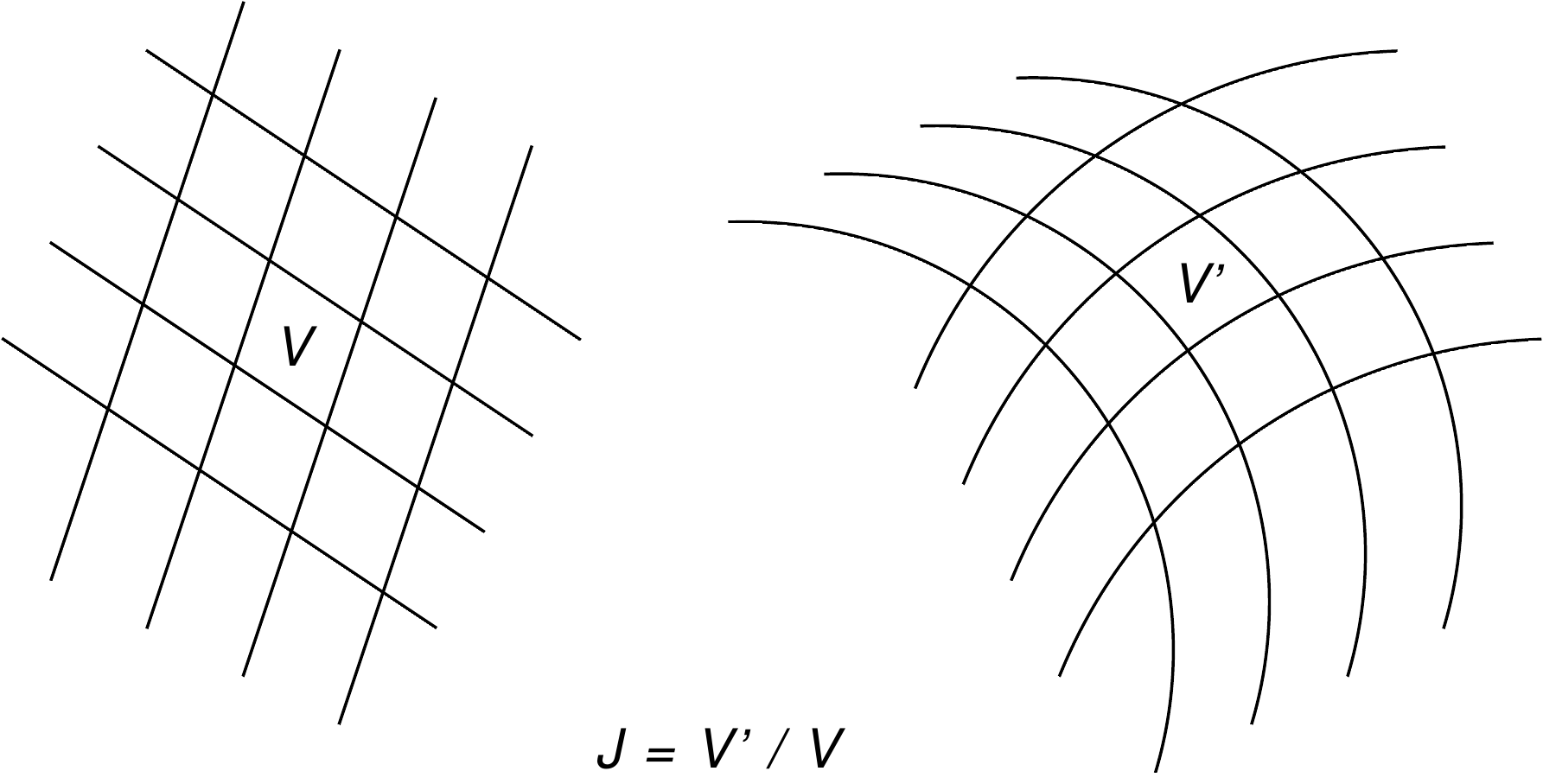}
\caption{The Jacobian factor can be thought as the ratio of an infinitesimal volume around the given point of interest in the two different bases $\vec{X}$ and $\vec{P}$ labeled by $V$ and $V^\prime$ respectively.~Lines on the left in the $\vec{X}$ basis correspond to lines on the right in the $\vec{P}$ basis.~The volume $V$ is translated into the volume $V'$ on the right.~The Jacobian factor the particular point is therefore $J = V' / V$.}
\label{fig:JacobiansGeometricalInterpretation}
\end{figure}
\begin{figure}
\centering
\includegraphics[width=0.30\textwidth]{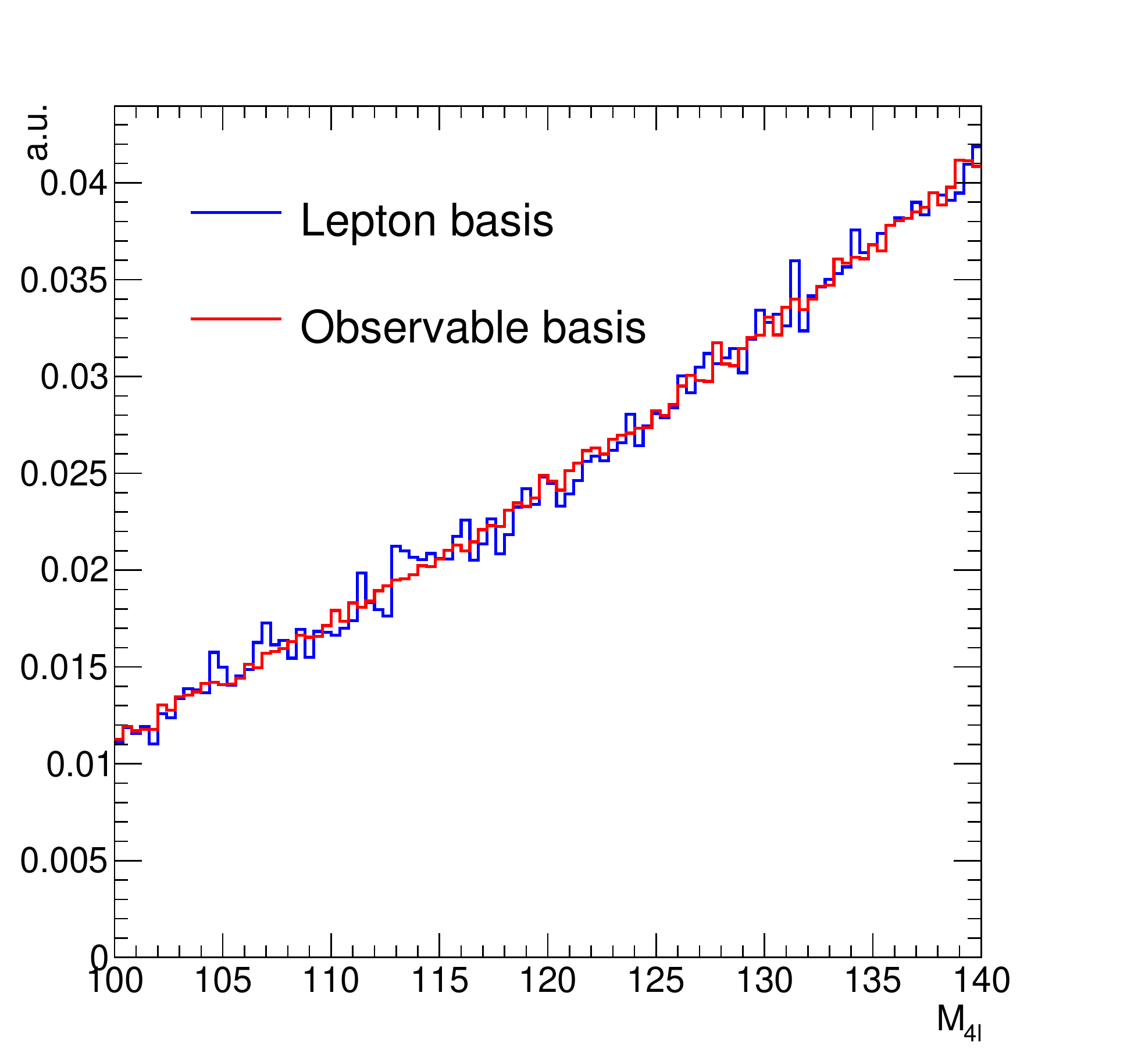}
\includegraphics[width=0.30\textwidth]{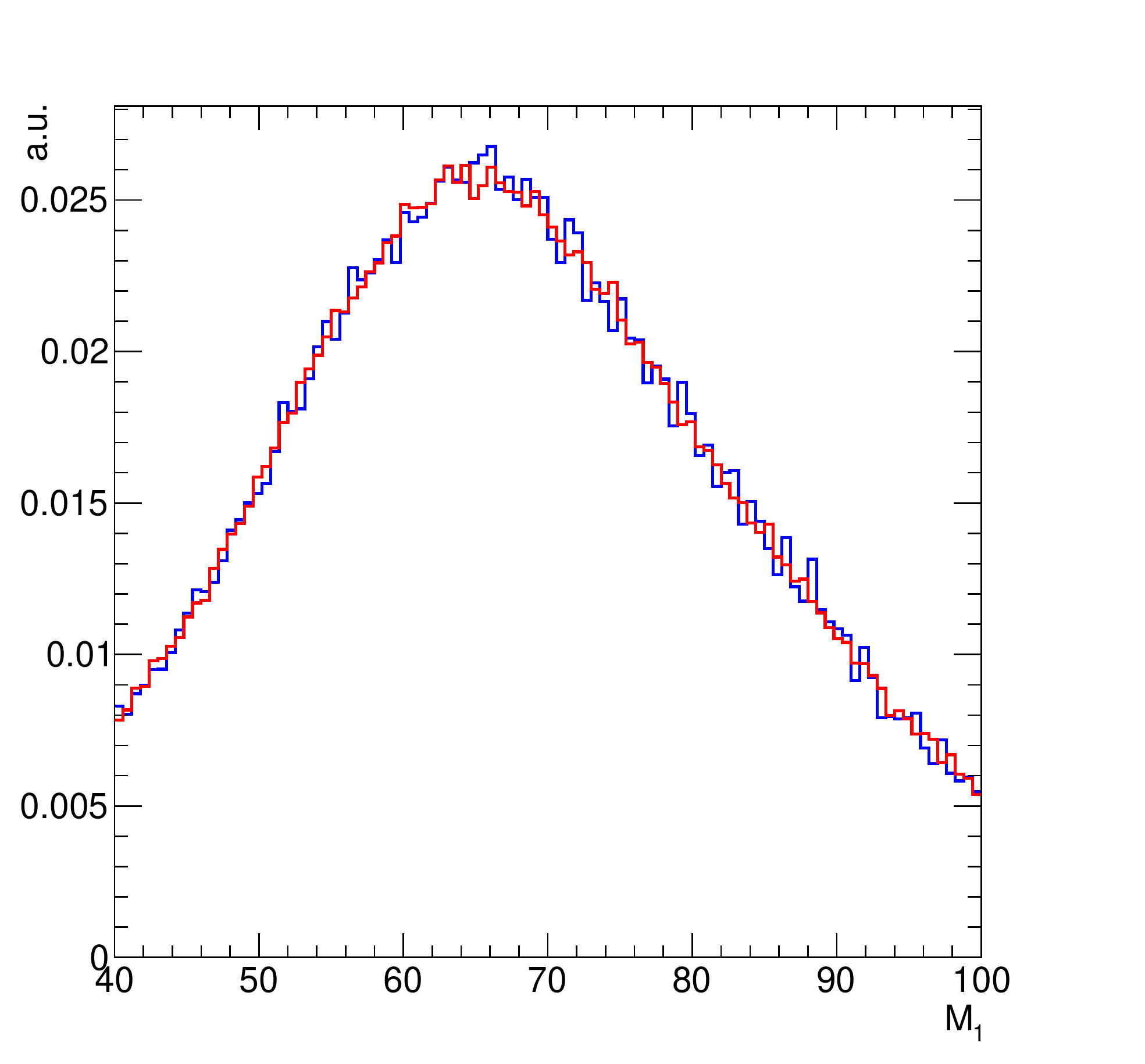}
\includegraphics[width=0.30\textwidth]{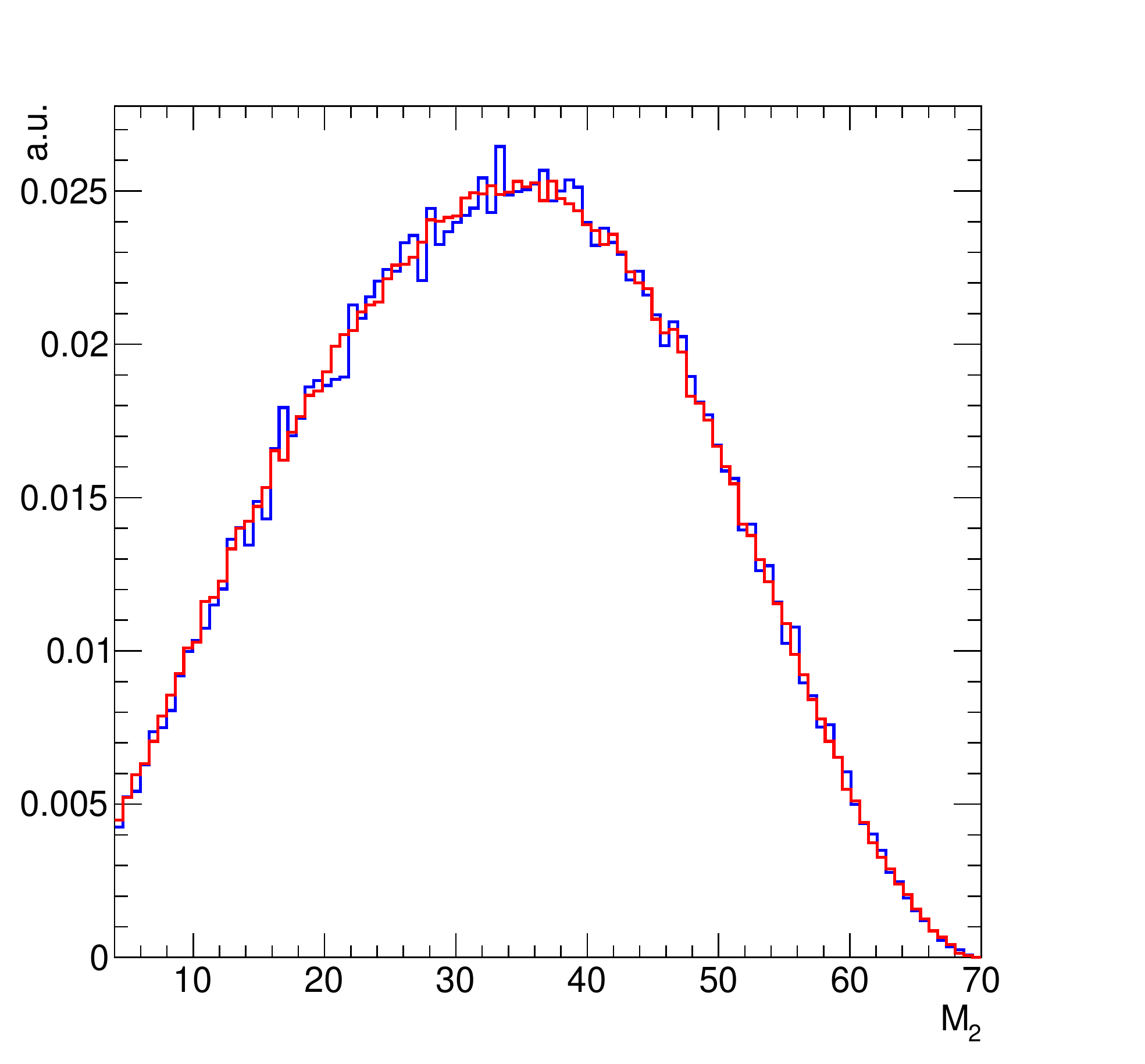}
\includegraphics[width=0.30\textwidth]{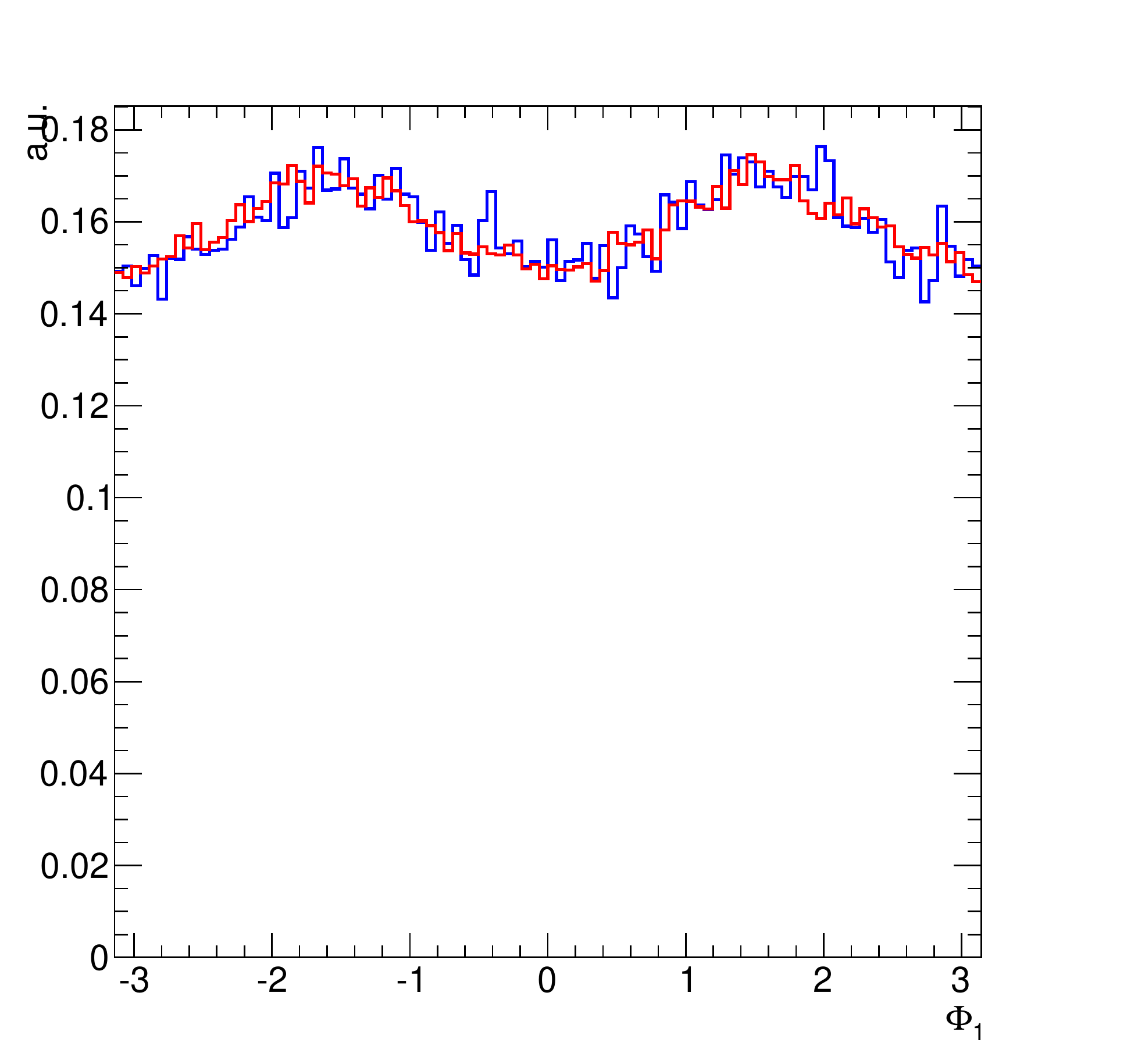}
\includegraphics[width=0.30\textwidth]{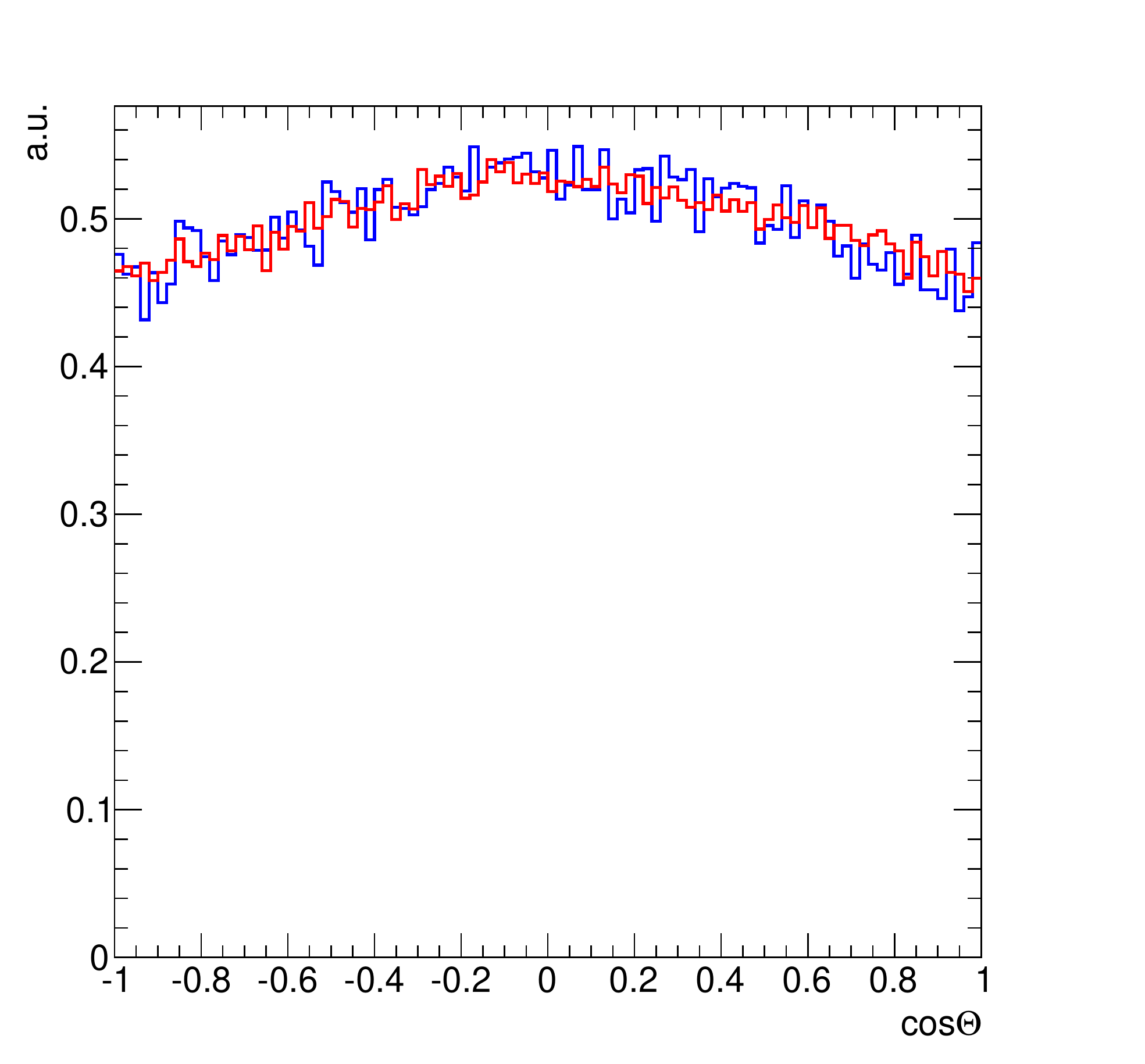}
\includegraphics[width=0.30\textwidth]{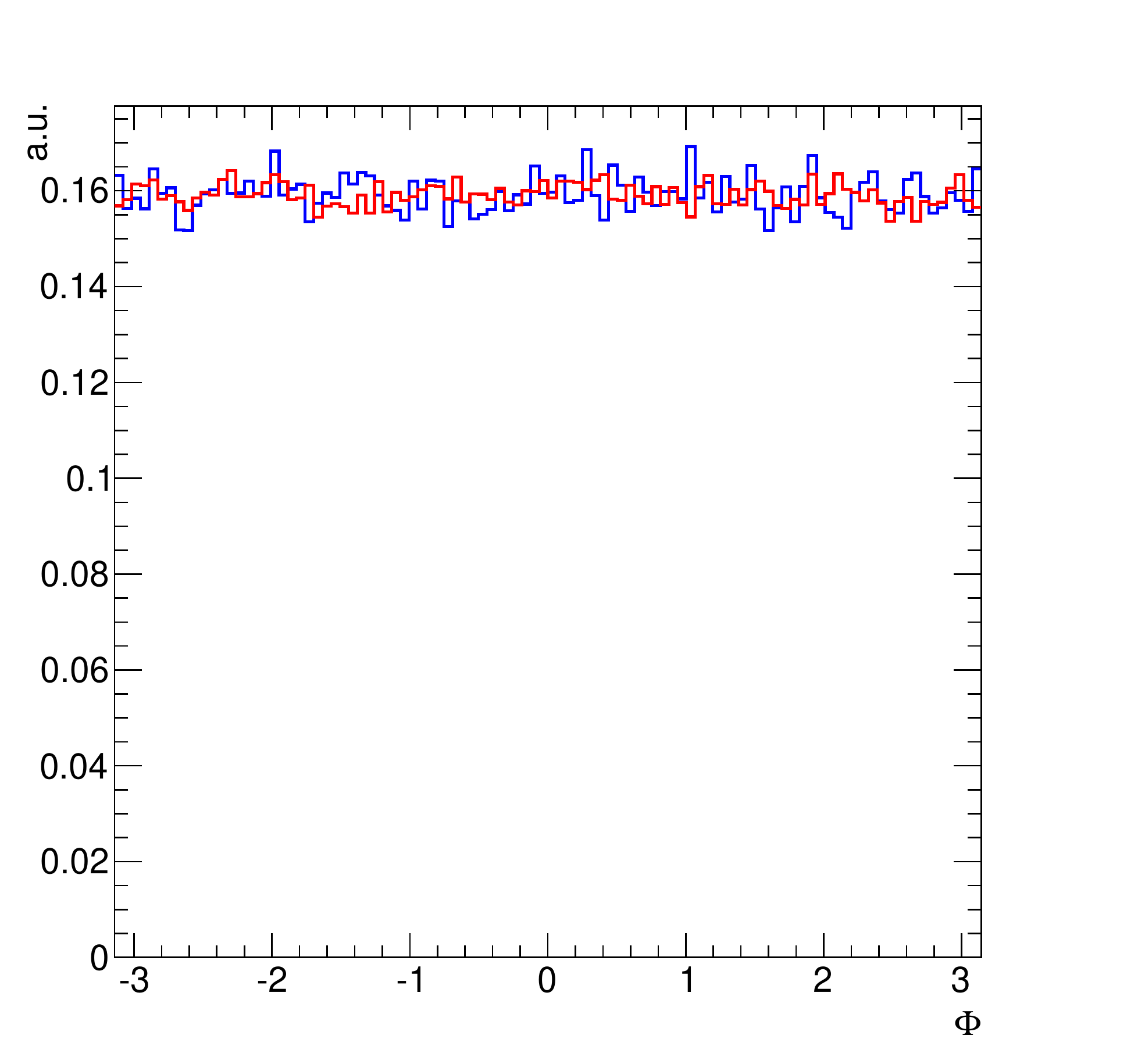}
\includegraphics[width=0.30\textwidth]{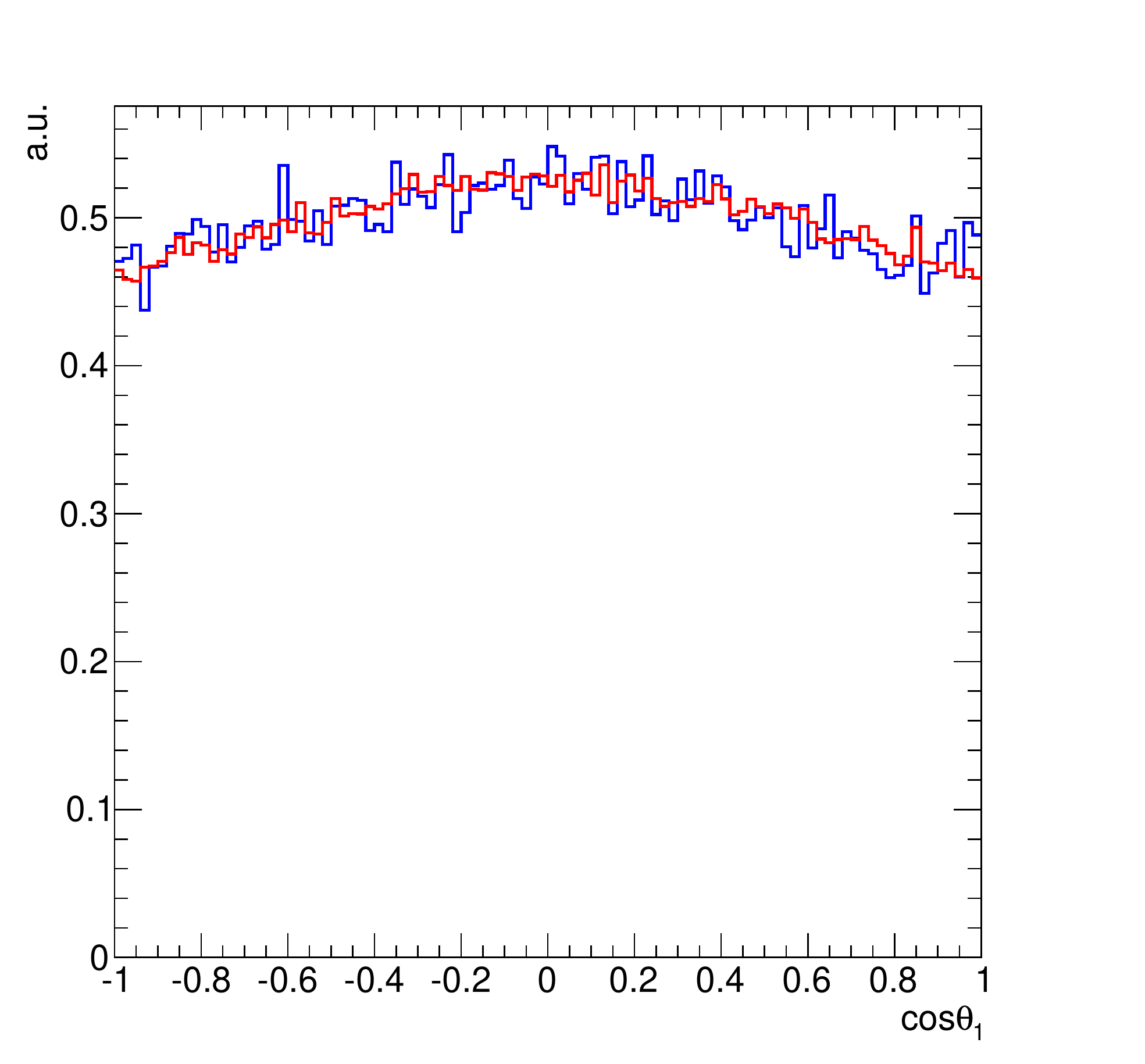}
\includegraphics[width=0.30\textwidth]{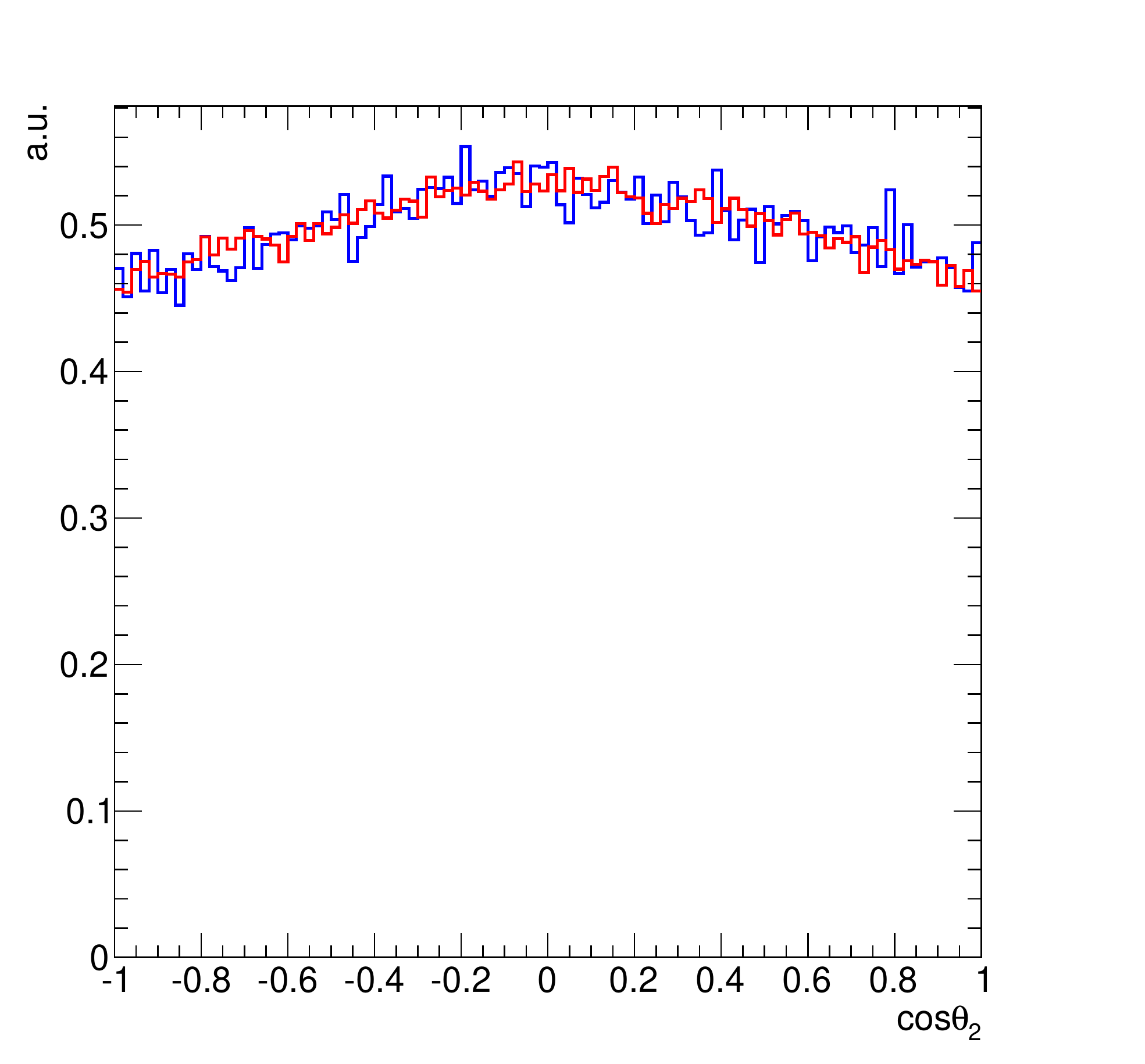}
\includegraphics[width=0.30\textwidth]{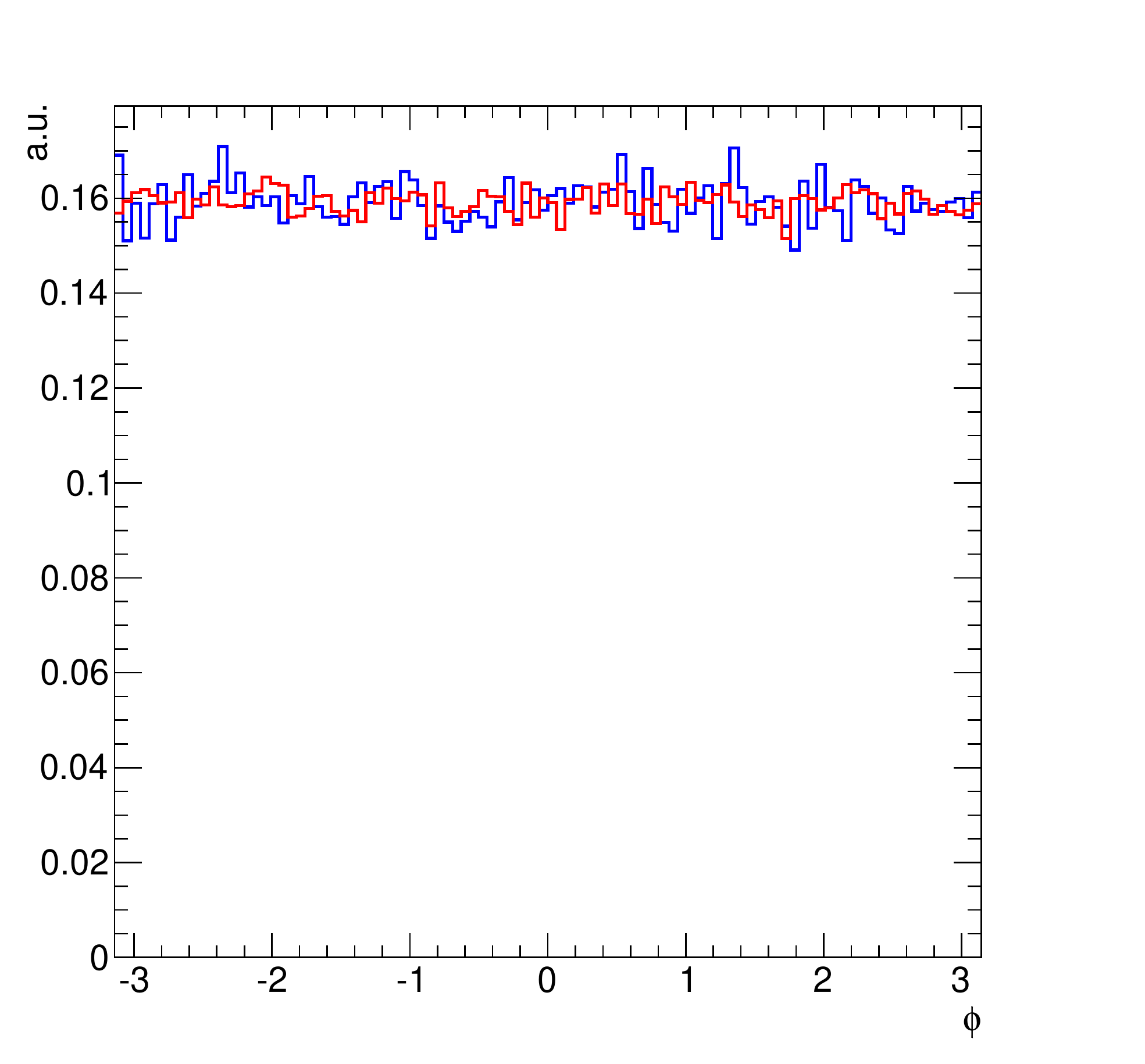}
\includegraphics[width=0.30\textwidth]{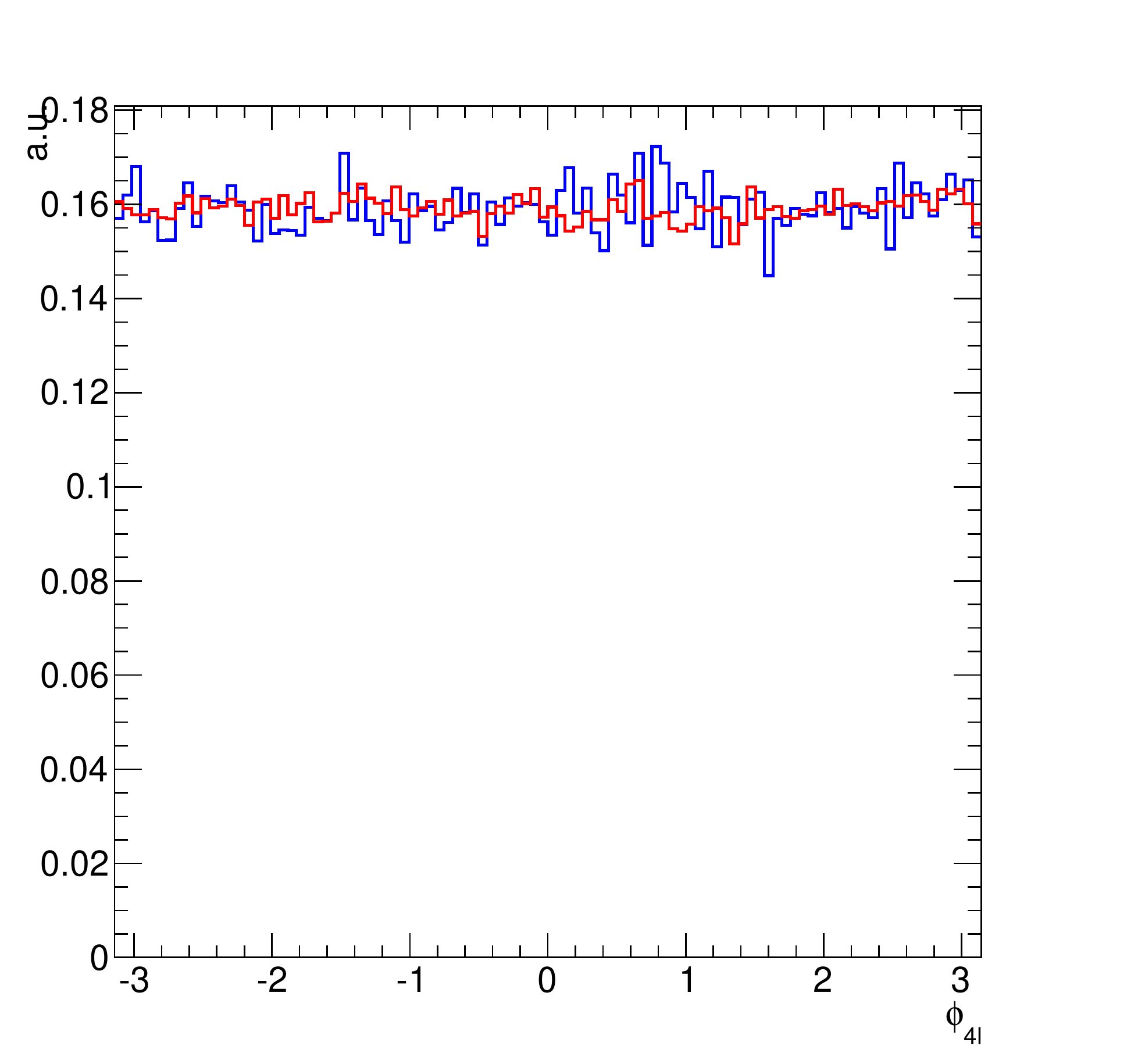}
\includegraphics[width=0.30\textwidth]{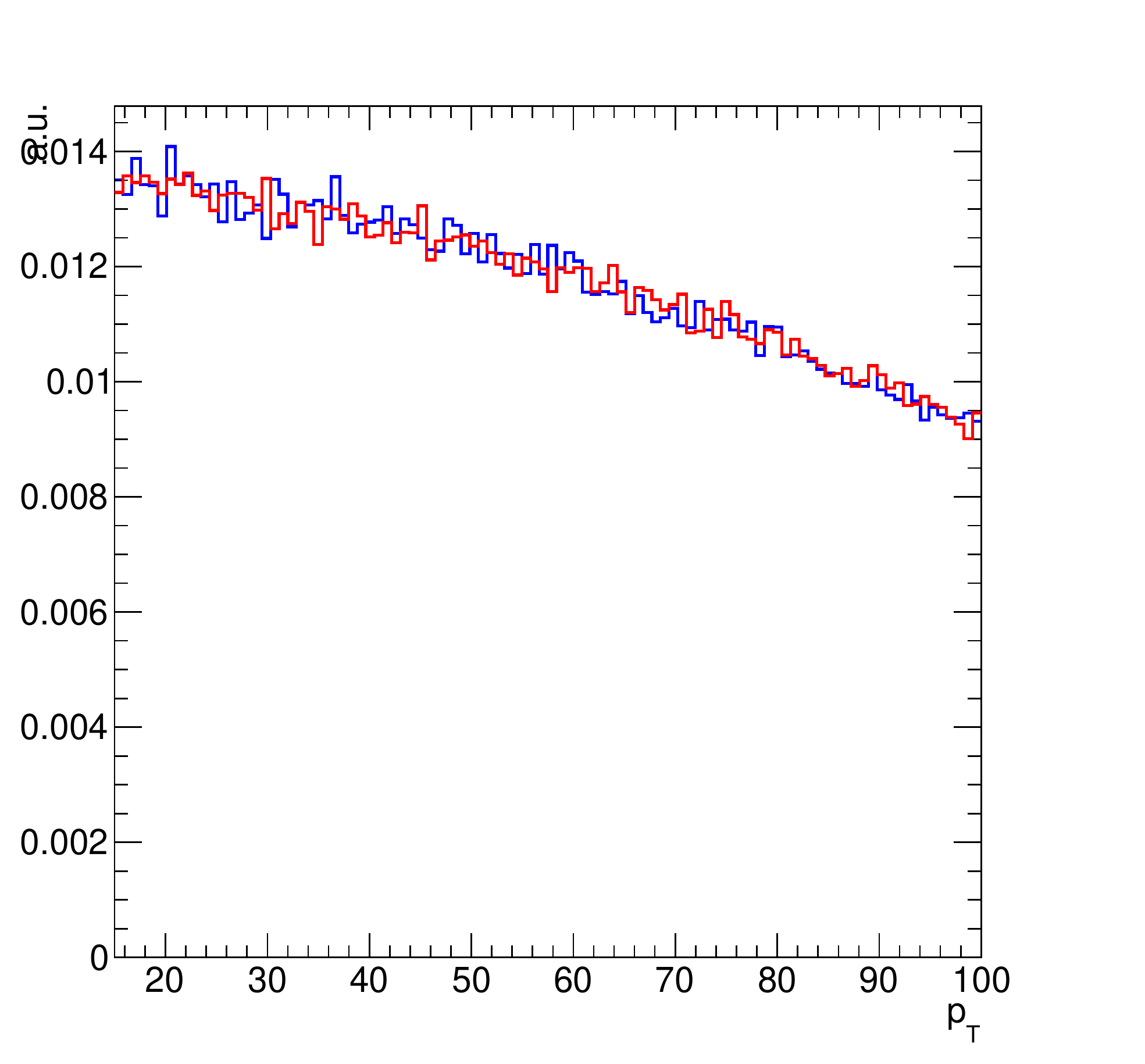}
\includegraphics[width=0.30\textwidth]{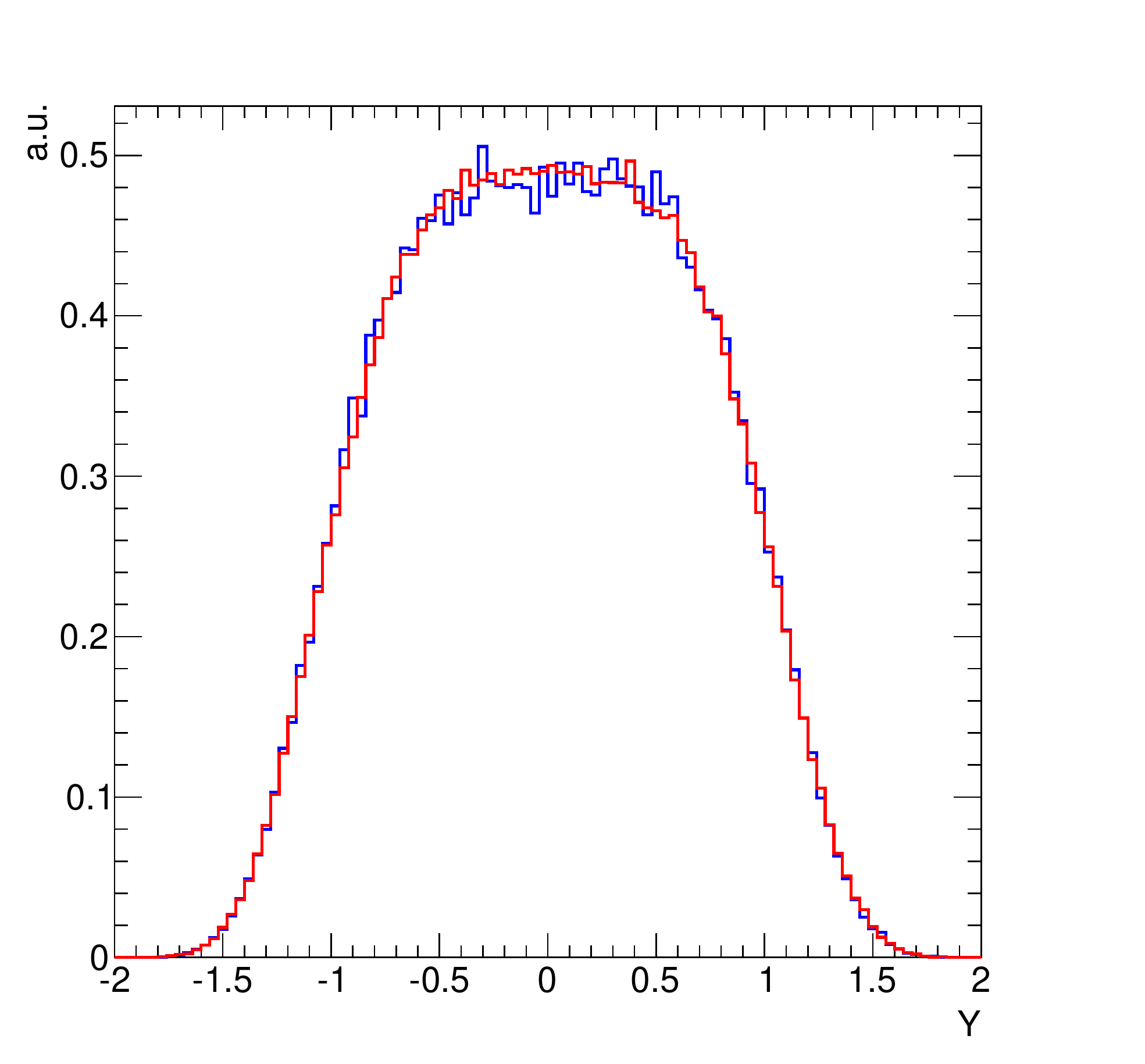}
\caption{Validation of the 12-dimensional Jacobian using the `toy' function in~\eref{Jcheck}.~Events are generated uniformly in both bases, and compared to each other by weighting those from the lepton basis $\vec{P}$ by their respective $12\times12$ Jacobian factor.~We see excellent agreement for all variables of interest with any differences due to statistical fluctuations.}
\label{fig:MultiDimensionalDetails-Jacobians-12DValidation}
\end{figure}

%%%%%%%%%%%%%%%%%%%%%%%%%%%%%%
%%%%%%%%%%%%%%%%%%%%%%%%%%%%%%
%%%%%%%%%%%%%%%%%%%%%%%%%%%%%%
\section{Integration Over Lepton Momenta}
\label{sec:MultiDimensionalDetailsIntegration}

We now turn to how the integration over the lepton momenta is performed in~\eref{reco_pdf3} and~\eref{reco_pdf_sig3}. The integral is performed by numerical methods based on Gaussian quadrature~\cite{Bradie:943117}.~We proceed by first scanning over a grid in the two di-lepton mass directions ($\mlla^2$ and $\mllb^2$)\footnote{In what follows we will no longer explicitly write the superscript $G$ for generator level observables which serve as our variables of integration.~We will however continue explicitly write the superscript $R$ for reconstructed level variables which are fixed quantities in the integration procedure.} and at each grid point integrating over the remaining $c_i$ smearing factors.~Over the grid of mass directions, it is natural that contributions to the final integral are concentrated around certain generator (`truth') level configurations.~A strategy is developed to maximize the precision of this procedure during the convolution integration which we now discuss.

%%%%%%%%%%%%%%%%%%%%%%%%%%%%%%
\subsection{Scan Directions}
\label{subsec:MultiDimensionalDetails-Integration-GridDirection}

We observe that in the background case the two mass directions are not correlated.~Variation in one mass direction is not dependent on the other mass direction.~Therefore a scan in the two mass directions (along $M_1$ and $M_2$) is a sufficiently good choice.~In the signal case, however, the correlation becomes much stronger due to the narrow width of the resonance and scanning along these directions is no longer optimal.~By constraining the mass of the four lepton system, a negative correlation is introduced to the two di-lepton masses.~It is thus advantageous to pick a ``diagonal direction'' as the direction of scanning over the grid.~This is done by defining `mass scan variables' as follow,
\bea
\mplus^2 &=& {\mlla^2} + \mllb^2 + R_m \mlla \mllb\nonumber,\\
\mminus^2 &=& (\mlla - \mllb)^2,
\eea
where $R_m$ is formed using \emph{reconstructed} level di-lepton masses as,
\begin{equation}
R_m \equiv \dfrac{{m_{13}^R}^2 + {m_{14}^R}^2 + {m_{23}^R}^2 + {m_{24}^R}^2}{{m_{12}^R} {m_{34}^R}},
\end{equation}
and ${m_{ij}^R}$ are the invariant masses formed by reconstructed lepton $i$ and lepton $j$.~Reconstructed masses are fixed and taken as input in the integration process.~By doing the scan in these directions, we obtain an additional Jacobian
factor which is easily calculated to be,
\bea
J_m = (R_m + 2)(\mlla^2 + \mllb^2).
\eea
An example of grid lines along this modified direction is shown in~\fref{MultiDimensionalDetails-Integration-MassScanDirection}.
\begin{figure}
\centering
\includegraphics[width=0.45\textwidth]{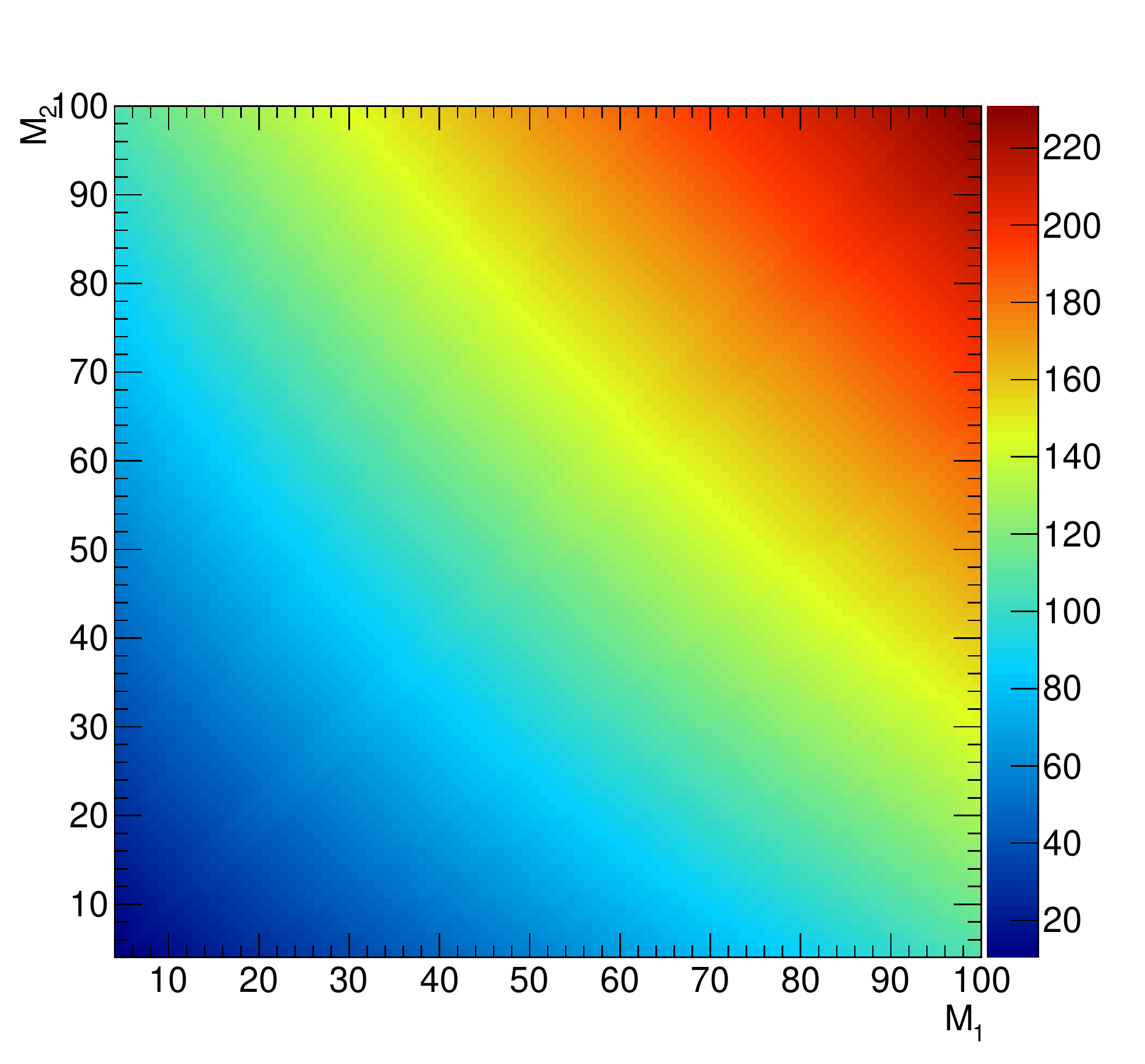}
\includegraphics[width=0.45\textwidth]{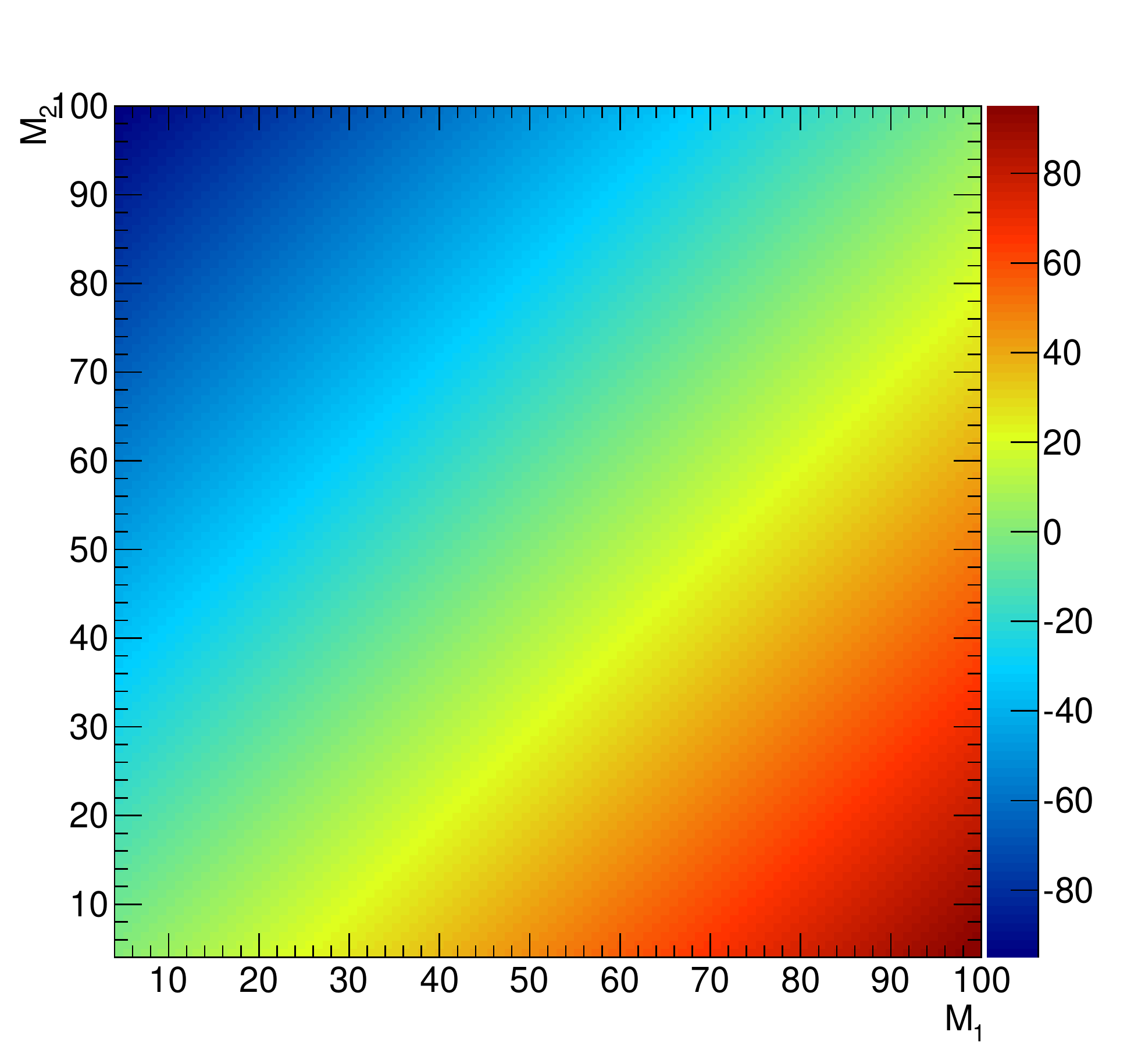}
\caption{Contour of \mplus (left) and \mminus (right) for one example event as a function
of $\mlla$ and $\mllb$.~The contour for \mplus changes from event to event, but is always
roughly diagonal.}
\label{fig:MultiDimensionalDetails-Integration-MassScanDirection}
\end{figure}

%%%%%%%%%%%%%%%%%%%%%%%%%%%%%
\subsection{Non-Uniform Grid Spacing}
\label{Subsection:MultiDimensionalDetails-Integration-GridSpacing}

In addition to using the mass scan directions discussed above, we also use a non-uniform grid for further optimization.~Due to the nature of the contributions concentrated around a certain `truth level' parameter point, we choose the grid to be more dense in the center.~For a uniform grid, one can write the location $\bar{x}_i$ of each grid point as,
\bea
\label{eqn:unifgp}
&& \bar{x}_i = x_o + \delta x_i, \nonumber \\
&& \delta x_i \equiv \dfrac{ \Delta x }{N_{\text{grid}}} \left(i - \dfrac{1}{2} N_{\text{grid}}\right) ,
\eea
where $x_o $ is the center point of the scan within the grid (which we attempt to place near the `true' point), $\Delta x$ is the distance between the leftmost point and rightmost point of the grid, and $N_{\text{grid}}$ is the number of grid points.

When allowing for non-uniform grid spacing,~\eref{unifgp} for the location of the grid point must be modified.~The grid spacing can be characterized by an ``attractor'' parameter $A_S$.~With this attractor we now define the new modified grid point locations $\tilde{x}_i$ as,
\bea
\tilde{x}_i = x_o+ \dfrac{\delta x_i(|A_S \delta x_i | + 1)}{ |A_S \Delta x |/2 + 1}.
\eea
The linear spacing is now modified to be quadratic, with center point and end points the same as before.~Larger $|A_S|$ values result in denser grid spacing near the center with linear spacing recovered when $A_S = 0$ giving $\tilde{x} \to \bar{x}$.~This behavior is illustrated in~\fref{MultiDimensionalDetails-Integration-Attractor}.~How the central point $x_o$ is chosen will be discussed in~\ssref{MultiDimensionalDetails-Integration-GridCenter}.
\begin{figure}
\centering
\includegraphics[width=0.8\textwidth]{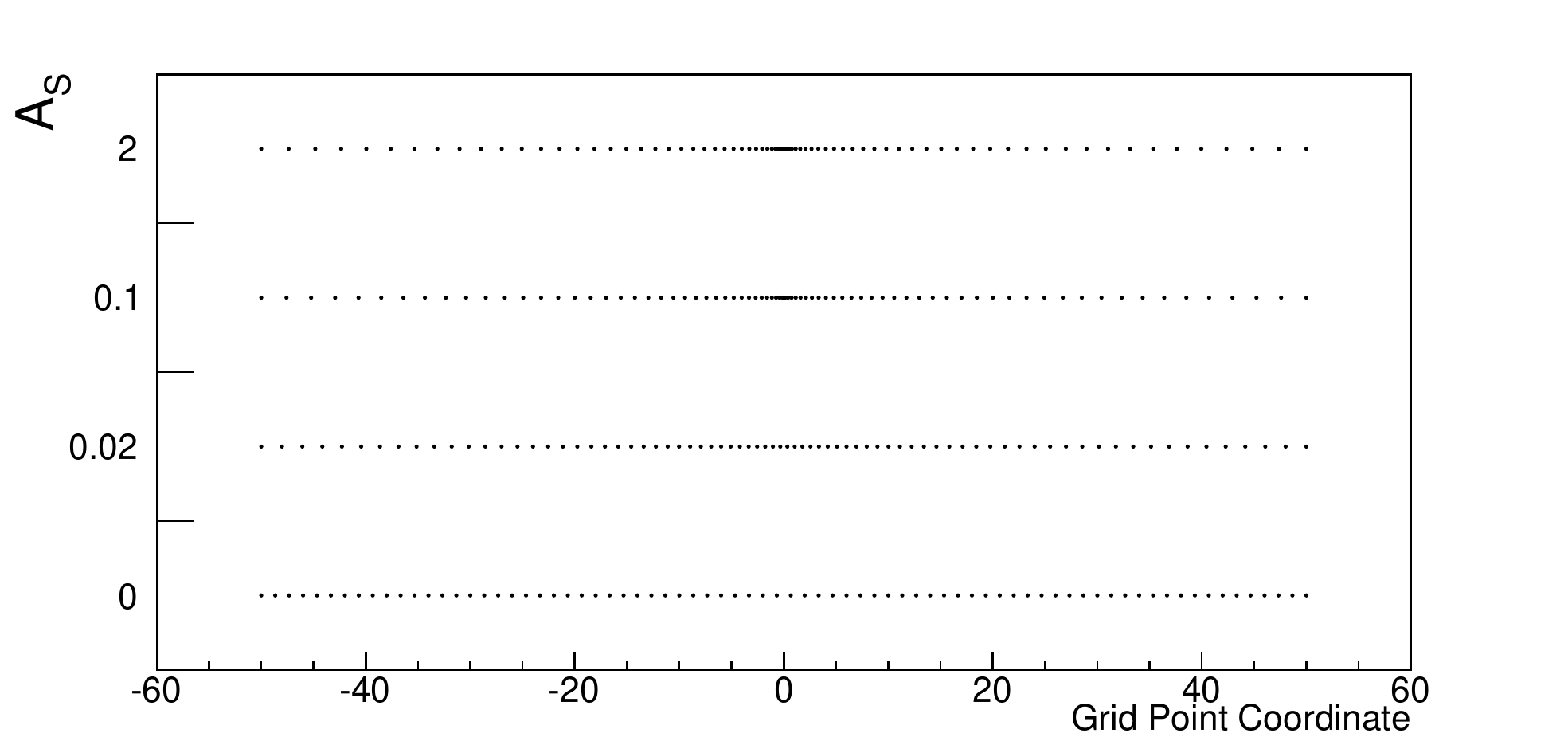}
\caption{Demonstration of the mass grid attractor.~Different sets of grid points are plotted in each line with varying mass grid attractor strength.~With attractor strength set to zero, we recover uniform spacing (bottom), while large values of attractor strength causes points to be concentrated near the center (top).}
\label{fig:MultiDimensionalDetails-Integration-Attractor}
\end{figure}
%

%%%%%%%%%%%%%%%%%%%%%%%%%%%%
\subsection{Modified 2nd Order Newton-Cotes Formula}
\label{Subsection:MultiDimensionalDetails-Integration-2ndOrderNewtonCotes}

With the modification of grid spacing,~it is necessary to derive the equivalent
of Newton-Cotes formula~\cite{Bradie:943117} for a non-uniform grid.~We work with 2nd order
closed integration where to each interval we assign three points to which we fit a second-order polynomial and then obtain the integral (area).~Furthermore, the area can be written as a weighted sum of the points used in the fit.~This can be seen by considering three points located at $-\delta, 0$, and $+\delta$ with height $f(-\delta), f(0)$ and $f(+\delta)$ respectively.~In the case of uniform gird spacing the area (of the function $f(x)$) estimator can then be written as the following~\cite{Bradie:943117},
\bea
\label{eqn:unifint}
I(-\delta, +\delta) = 2\delta \, \dfrac{f(-\delta) + 4 f(0) + f(\delta)}{6}.
\label{eqn:MultiDimensionalDetails-Integration-LinearIntegral}
\eea
For non-uniform grid spacings,~\eref{unifint} needs to be re-derived.~Since the integral does not depend on the absolute value of the x-axis coordinates, we can take the liberty to pick the center point at zero, and the other two located at $x = -\delta_{-}$ and $x = \delta_+$ for the left and right endpoints of the interval respectively.~Integrands at the different points can be written as $f(-\delta_-), f(0), f(\delta_+)$.
For non-uniform spacing the integral estimator is modified to,
\begin{align}
I(-\delta_-, \delta_+) &= \dfrac{1}{3} A (\delta_+^3 + \delta_-^3)
 + \dfrac{1}{2} B (\delta_+^2 - \delta_-^2) + C (\delta_+ + \delta_-),
\label{eqn:MultiDimensionalDetails-Integration-2ndOrderNewtonCotesIntegral}
\end{align}
where $A, B$ and $C$ are the coefficients of the 2'nd order polynomial which is used in the fit to the points in a given interval and defined as,
\bea
y(x) = Ax^2 + Bx + C.
\eea
In terms of the three points in the interval the coefficients are obtained via the matrix equation,
\begin{align}
\left[ \begin{array}{c}
A \\ B \\ C
\end{array} \right] = 
\dfrac{1}{\delta_+\delta_- (\delta_+ + \delta_-)}
\left[\begin{array}{ccc}
\delta_- & -\delta_+ - \delta_- & \delta_+\\
\delta_-^2 & \delta_+^2 - \delta_-^2 & -\delta_+^2\\
0 & \delta_+ \delta_- (\delta_+ + \delta_-) & 0
\end{array}\right]
\left[\begin{array}{c}
f(\delta_+) \\ f(0) \\ f(\delta_-)
\end{array}\right].
\label{eqn:MultiDimensionalDetails-Integration-2ndOrderNewtonCotesCoefficients}
\end{align}
Together~\eref{MultiDimensionalDetails-Integration-2ndOrderNewtonCotesIntegral} and~\eref{MultiDimensionalDetails-Integration-2ndOrderNewtonCotesCoefficients} define the final reweighing coefficients to use for the non-uniform gird integration.~When $\delta_- = \delta_+$ they reduce to the linear formula in~\eref{MultiDimensionalDetails-Integration-LinearIntegral}.~The analogous two-dimensional formula can be trivially obtained by using~\eref{MultiDimensionalDetails-Integration-2ndOrderNewtonCotesIntegral} and~\eref{MultiDimensionalDetails-Integration-2ndOrderNewtonCotesCoefficients} multiplicatively on the two dimensions.

%%%%%%%%%%%%%%%%%%%%%%%%%%%%
\subsection{Central Grid Point Optimization}
\label{subsec:MultiDimensionalDetails-Integration-GridCenter}

During the mass scan, we do not know \emph{a priori} where the location of the maximum contribution to the final integral will be.~If we do not center our mass grid scan close to the maximum contribution point, there is a chance part of the contributions will be missed.~It is especially crucial when we consider non-uniform grid spacings to increase efficiency in mass integration.

We therefore employ a simple numerical algorithm to find where the maximum contribution
point is.~It is outlined as follows:
\begin{itemize}
\item For the background we start from a best-guess value of reconstructed masses $M_1^R$ and $M_2^R$ while choosing a reasonable window size.
\item For signal we choose the best-guess value to be $(\sqrt{\hat{s}/\hat{s}^R})M_1^R$ and $(\sqrt{\hat{s}/\hat{s}^R})M_2^R$ again choosing a reasonable window size.
\item We put a coarse grid in this window centered at the current best guess value and evaluate
the integrand (integrals of $c_1$ and $c_3$) at each point.
\item The integration is carried out repeatedly with the central point updated each time to the point which gives the largest contribution.
\item We then reduce the window size by a fraction and repeat the process.
\item The process terminates when the point with the largest contribution lies within the $5\%$ most central grid points after which we adjust the grid window.
\item The final best guess value is used as the center point for constructing the mass grid.
\end{itemize}
During the integration we also keep track of the RMS of the integral contribution in units of number of grid points.~If the RMS is found to be less than four grid points in either direction,~the integration is repeated with a reduced window.~This ensures us of a grid center that is sufficiently close to the `truth level' point with maximum contribution.

A demonstration of the whole process on one example signal event is shown in~\fref{MultiDimensionalDetails-Integration-AttractorExample}.~Each bin in the plot is one grid point which has been scanned over and the color indicates the inner integral from that given bin.~On the left we have the usual scanning directions (along $M_1$ and $M_2$) and on the right we show the scan along the modified diagonal directions (along $m_+$ and $m_-$) as explained in~\ssref{MultiDimensionalDetails-Integration-GridDirection}.~In the top two panels the scanning is done without implementing the mass grid attractor and without the central grid point optimization.~In the middle two panels the mass grid attractor is turned on, but without central grid point optimization.~In the bottom panels both the mass grid attractor and central grid point optimizations are turned on.~With all optimizations `turned on' the amount of grid points with significant contributions is greatly enhanced with decreasing correlation between the two scanning directions.~This provides us increasing precision and stability during the integration.
\begin{figure}
\centering
\includegraphics[width=0.43\textwidth]{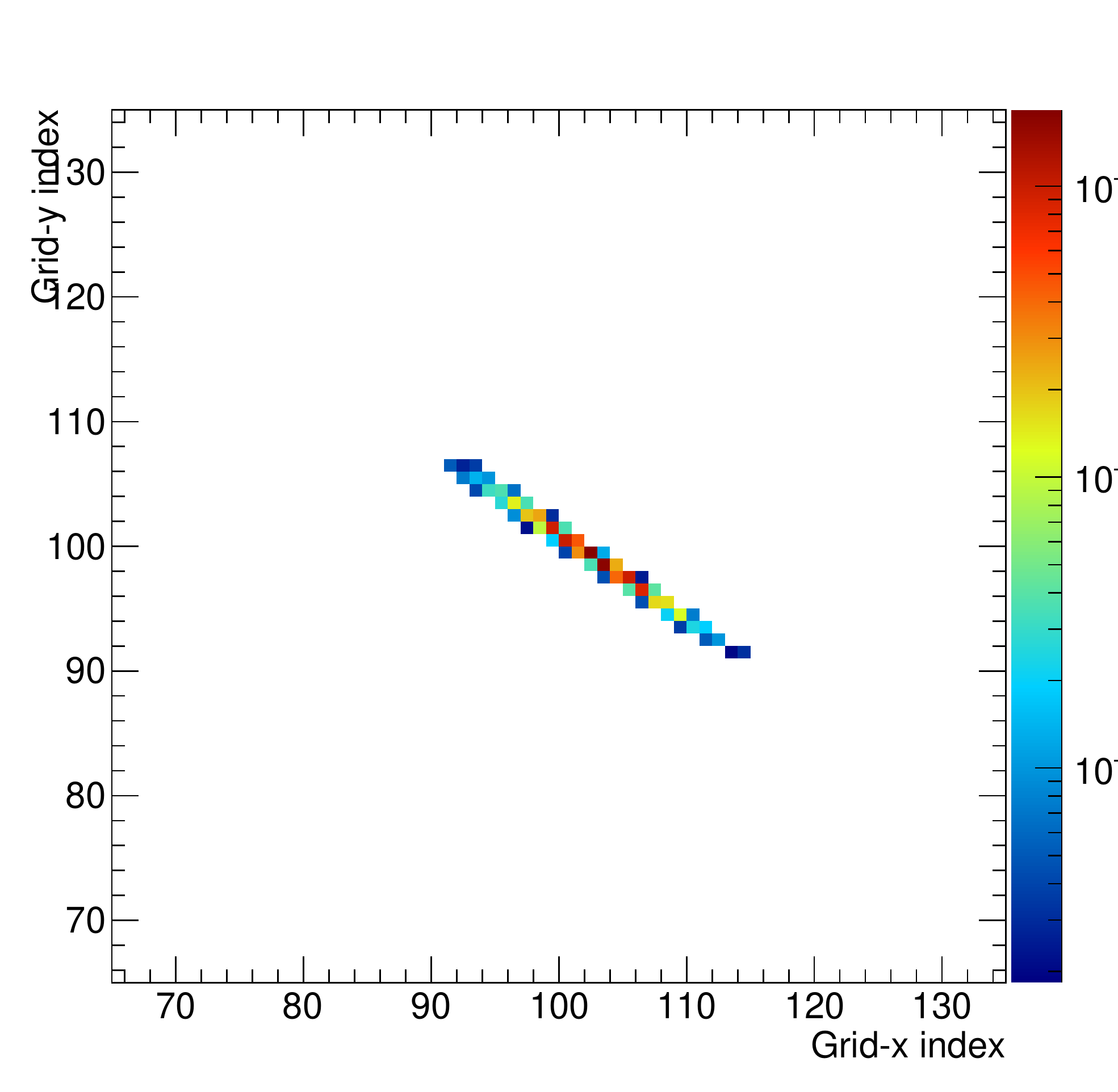}
\includegraphics[width=0.43\textwidth]{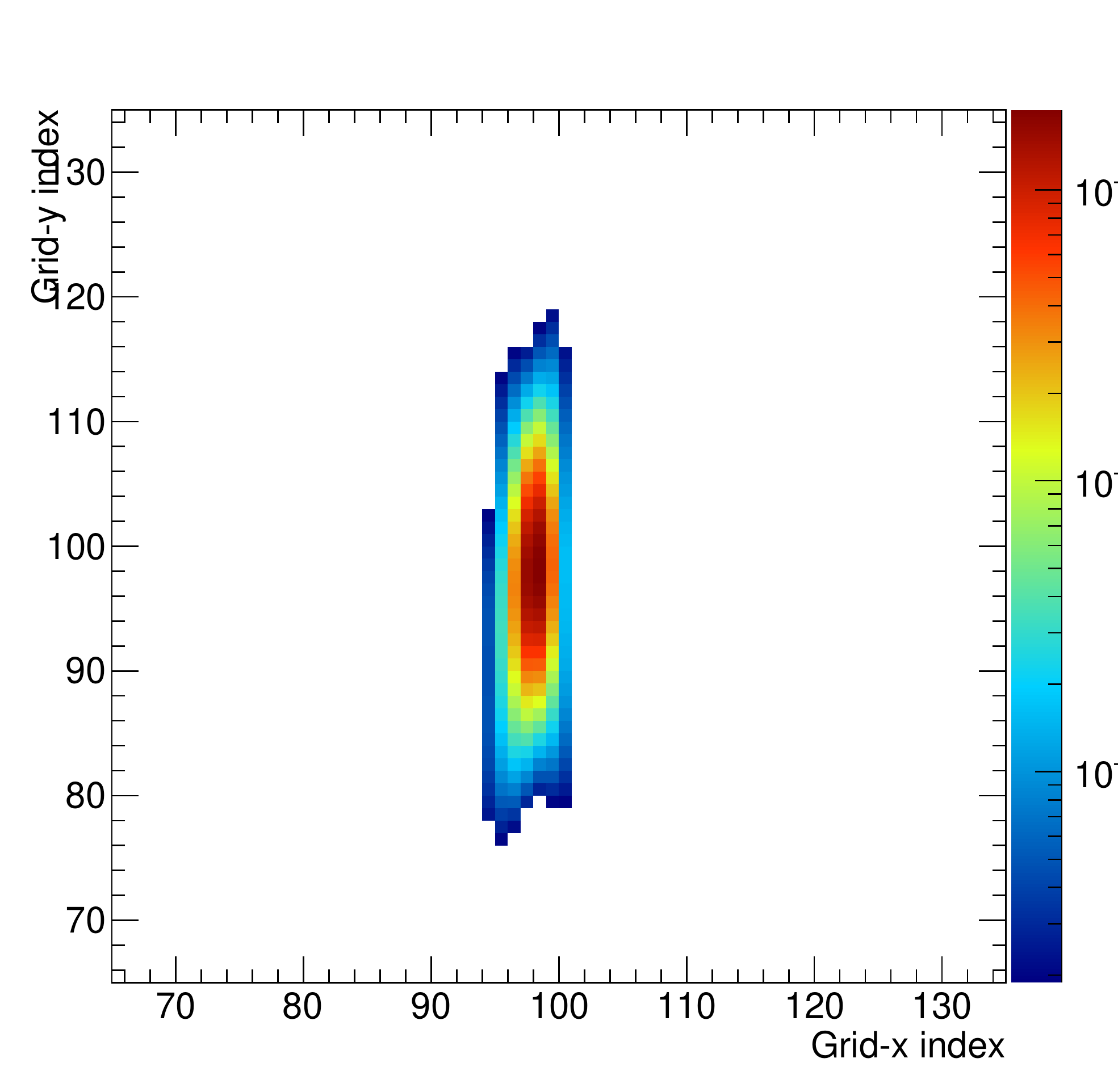}
\includegraphics[width=0.43\textwidth]{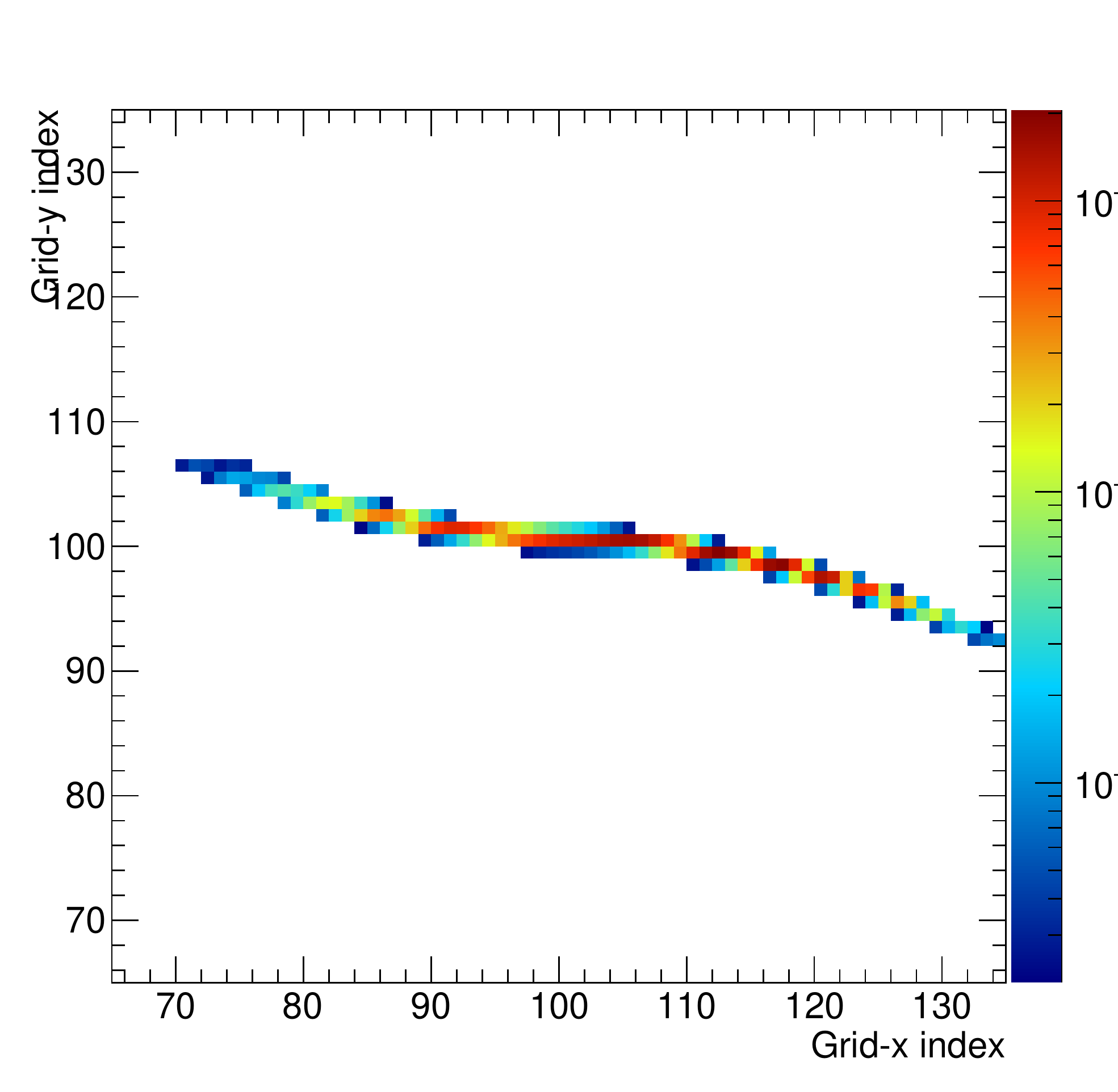}
\includegraphics[width=0.43\textwidth]{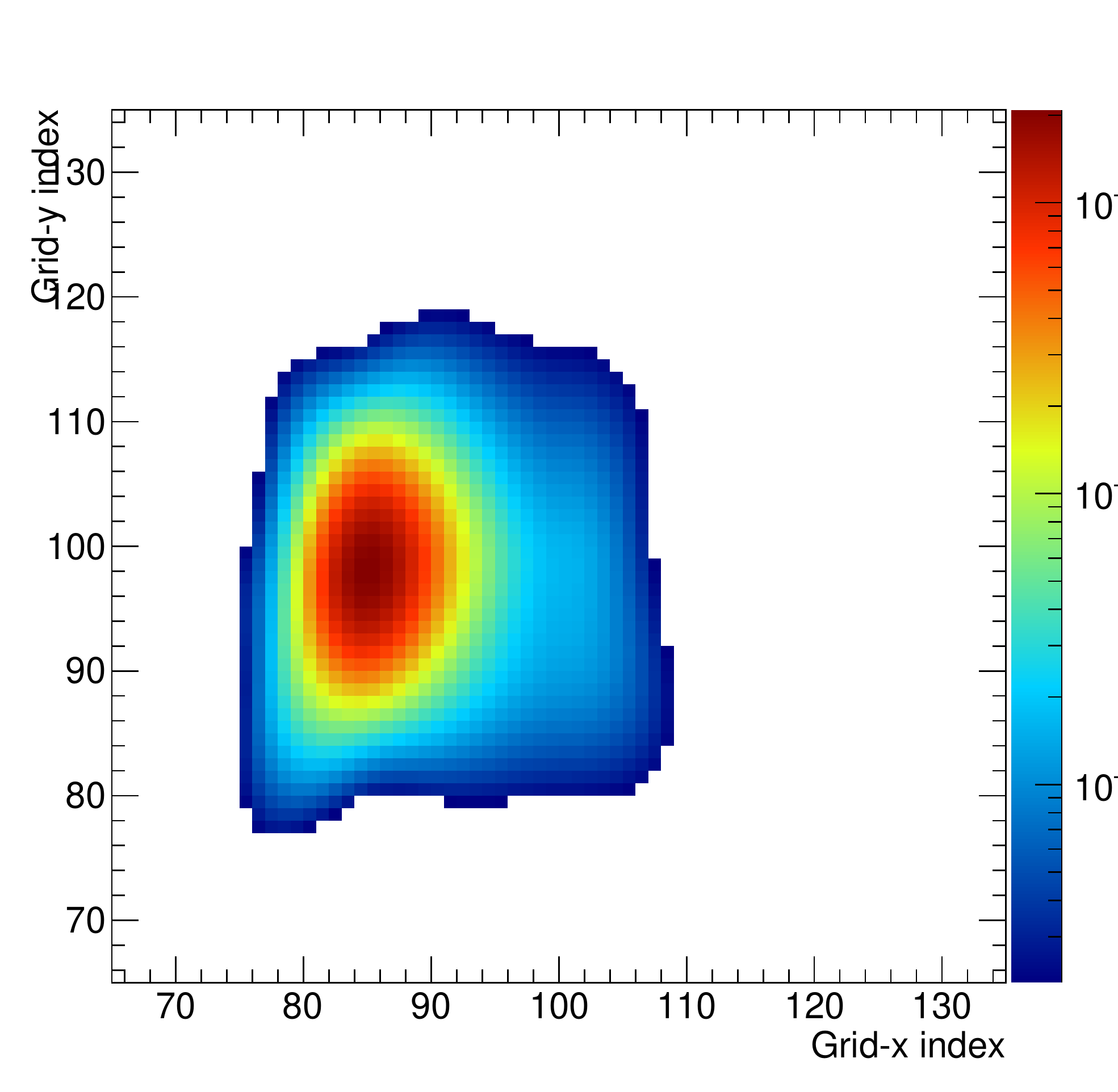}
\includegraphics[width=0.43\textwidth]{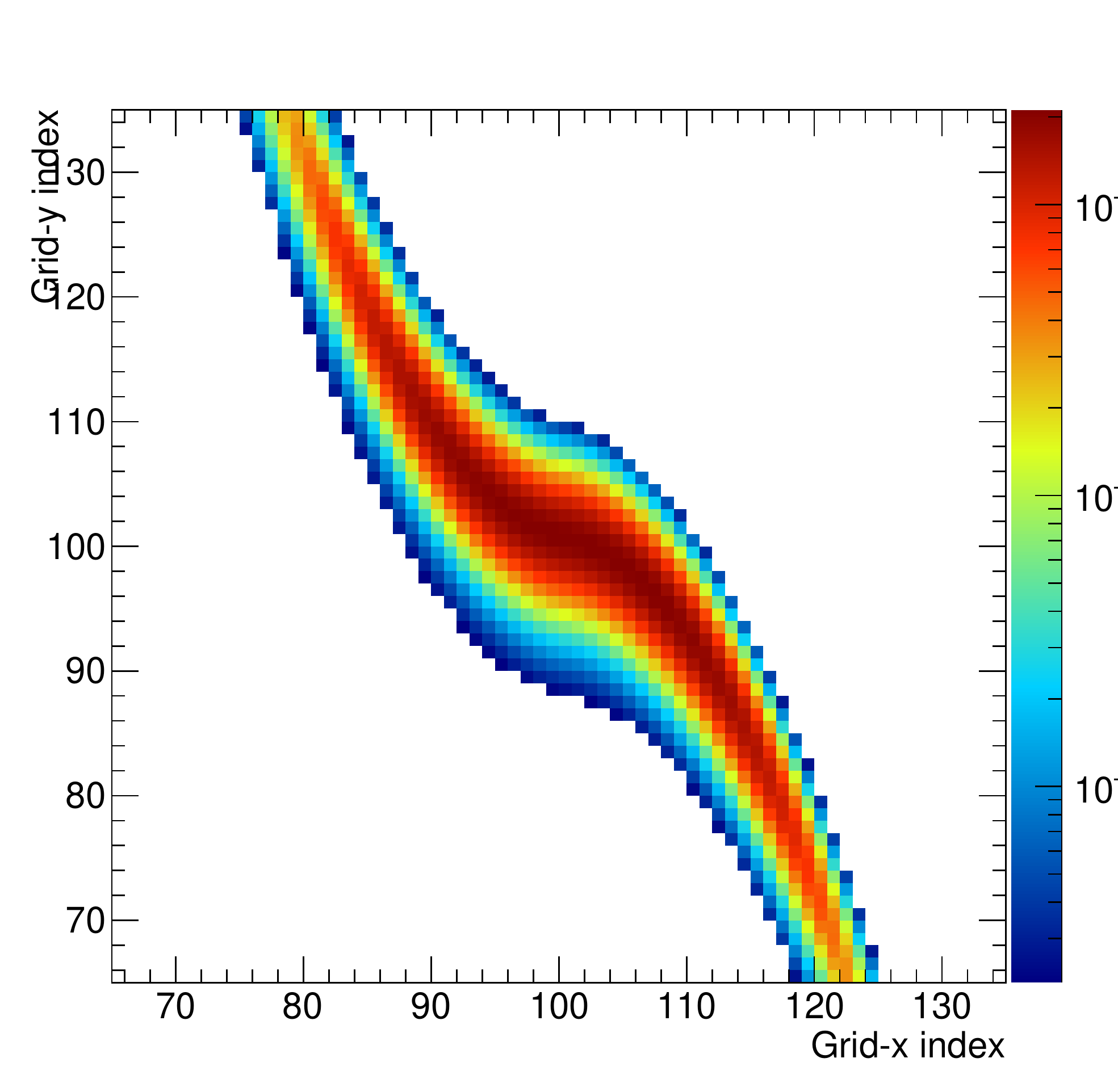}
\includegraphics[width=0.43\textwidth]{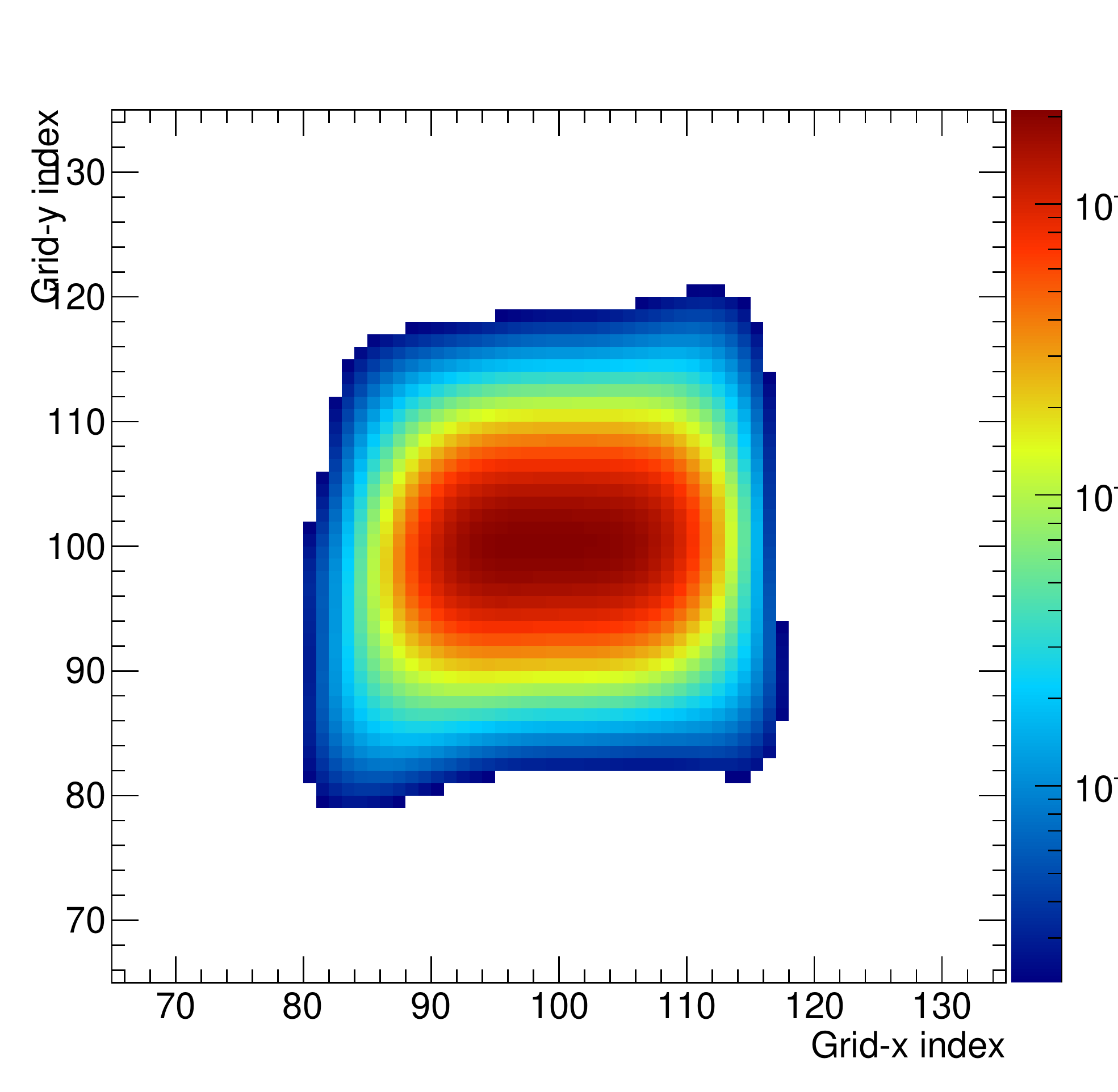}
\caption{Demonstration of mass grid attractor and scan directions during the integration.~Each bin is one grid point, where the color indicates the value of the inner integral.~In the top two plots the attractor strength is set to zero.~In the middle plots we have chosen a moderate attractor strength while in the bottom plots the center grid point optimization is turned on.~Plots on the left are along the scanning directions $M_1, M_2$ while on the right we show the modified directions \mplus, \mminus.}
\label{fig:MultiDimensionalDetails-Integration-AttractorExample}
\end{figure}
%

%%%%%%%%%%%%%%%%%%%%%%%%%%%%
\subsection{Differential Cross Section Expansion for Background}
\label{subsec:MultiDimensionDetails-Integration-Expansion}

We now turn to the integration over the smearing factors ($c_i$) in~\eref{reco_pdf3} and~\eref{reco_pdf_sig3}.~As will be discussed in more detail in the next section, we wish to further control precision by applying an adaptive integration method for the integrals over these variables.~In the background case, at each grid point for the masses, there is a two-dimensional integral over $c_1$ and $c_3$ which must be performed.~Since a brute force recursive integration method in two dimensions would take too much computing resources, we start by making a few observations.

We first note that in the final integration expression for the background (\eref{reco_pdf3}), the transfer function can be factorized into terms for the first lepton pair and terms for second lepton pair.~This assumes the transfer function is a product of functions for each lepton individually which is a very good approximation for the CMS and ATLAS detectors~\cite{CMSperformance2,CMS-DP-2013-003} once standard lepton selection criteria are imposed\footnote{If the approximation of sufficiently separated leptons does not hold, the integration procedure can not be carried out as formulated and must be altered.~We do not explore these complications associated with leptons with very small separations and can avoid them by simply imposing a minimum $m_{\ell\ell}$ cut between any two leptons.}.~Furthermore, the range of smearing factors is considered to be narrow since the probability that the lepton momentum is smeared far from its `true' value is very small~\cite{CMSperformance2,CMS-DP-2013-003,Thesis}.~This allows us to write the rest of the integrand, apart from the transfer function, as a series expansion in smearing parameters (about the `true' value of one) as follows,
\bea
\mathcal{F}_B(\vec{X}^R) 
&\equiv& 
P_B(\vec{X}^G) 
|{\bf J}^{\vec{P}}_G|
\frac{|{\bf J}^{\vec{c}}_G|}{|{\bf J}^{\vec{c}}_R|} 
|{\bf J}^{\vec{M}}_B| \nonumber \\
&\approx& \mathcal{F}_B\big|_{c_1=c_3=1} + \dfrac{\partial \mathcal{F}_B}{\partial c_1} 
\bigg|_{c_1=c_3=1} (c_1 - 1) + ... \nonumber\\
&\equiv& B_0 + B_1 (c_1 - 1) + B_3 (c_3 - 1) + B_{11} (c_1 - 1)^2\nonumber\\
&+& B_{13} (c_1 - 1) (c_3 - 1) + B_{33} (c_3 - 1)^2 + ...,
\eea
where $c_i = 1$ corresponds to the `truth level' value for the smearing factors and the differential cross sections are implicitly absorbed into the $B_{ij}$ coefficients.~We can then factorize the inner integral into a sum of products of single integrals.~Thus we have,
\bea
P_B(\vec{X}^R) &=& 
\dfrac{1}{|{\bf J}^{\vec{P}}_R|} \int P_B(\vec{X}^G) T(\vec{c} |\vec{P}^G) 
|{\bf J}^{\vec{P}}_G| \frac{|{\bf J}^{\vec{c}}_G|}{|{\bf J}^{\vec{c}}_R|} 
|{\bf J}^{\vec{M}}_B| 
dc_1 dc_3 d\mlla^2 d\mllb^2\nonumber\\
&=& \dfrac{1}{|{\bf J}^{\vec{P}}_R|} \int \left(P_B(\vec{X}^G) T(\vec{c} | \vec{P}^G) 
|{\bf J}^{\vec{P}}_G|
\frac{|{\bf J}^{\vec{c}}_G|}{|{\bf J}^{\vec{c}}_R|} 
|{\bf J}^{\vec{M}}_B| 
dc_1 dc_3 \right) d\mlla^2 d\mllb^2,\nonumber
\eea
where the inner integral can now be expanded as,
\bea
&&\int P_B(\vec{X}^G) T(\vec{c} | \vec{P}^G) 
|{\bf J}^{\vec{P}}_G|
\frac{|{\bf J}^{\vec{c}}_G|}{|{\bf J}^{\vec{c}}_R|} 
|{\bf J}^{\vec{M}}_B| 
dc_1 dc_3 \nonumber\\
&\approx& B_0 \int T_{12}(\vec{c} | \vec{P}^G) (c_1 - 1)^0 dc_1 \int T_{34}(\vec{c} | \vec{P}^G) (c_3 - 1)^0 dc_3 \nonumber\\
&+& B_1 \int T_{12}(\vec{c} | \vec{P}^G) (c_1 - 1)^1 dc_1 \int T_{34}(\vec{c} | \vec{P}^G) (c_3 - 1)^0 dc_3 \nonumber\\ 
&+& B_3 \int T_{12}(\vec{c} | \vec{P}^G) (c_1 - 1)^0 dc_1 \int T_{34}(\vec{c} | \vec{P}^G) (c_3 - 1)^1 dc_3 \nonumber\\ 
&+& B_{11} \int T_{12}(\vec{c} | \vec{P}^G) (c_1 - 1)^2 dc_1 \int T_{34}(\vec{c} | \vec{P}^G) (c_3 - 1)^0 dc_3 \nonumber\\ 
&+& B_{13} \int T_{12}(\vec{c} | \vec{P}^G) (c_1 - 1)^1 dc_1 \int T_{34}(\vec{c} | \vec{P}^G) (c_3 - 1)^1 dc_3 \nonumber\\ 
&+& B_{33} \int T_{12}(\vec{c} | \vec{P}^G) (c_1 - 1)^0 dc_1 \int T_{34}(\vec{c} | \vec{P}^G) (c_3 - 1)^2 dc_3 
+ ...\nonumber\\
&\equiv& B_0 F_{12}^{(0)} F_{34}^{(0)} + B_1 F_{12}^{(1)} F_{34}^{(0)} + B_3 F_{12}^{(0)} F_{34}^{(1)}\nonumber\\
&+& B_{11} F_{12}^{(2)} F_{34}^{(0)} + B_{13} F_{12}^{(1)} F_{34}^{(1)} + B_{33} F_{12}^{(0)} F_{34}^{(2)} + ...~,
\eea
and we have defined the integrals,
\bea
F_{ij}^{(n)} \equiv \int T_{ij}(\vec{c} | \vec{P}^G) (c_i - 1)^n d c_i .
\eea
The transfer functions for the first and second lepton pairs are denoted as $T_{12}(\vec{c} | \vec{P}^G)$ and $T_{34}(\vec{c} | \vec{P}^G)$ respectively.~Thus we see that the two-dimensional integral over the smearing parameters has been reduced to a product of single integrals which are computationally easier to control.~With this procedure one can increase the expansion order if higher precision is required.~The $B_{ij}$ coefficients in front of each $F_{ij}^{(n)}$ term can be obtained by finding an approximate two-dimensional polynomial to the target function over the whole integration range.~In the current implementation it is done by picking a few points as representative and finding a polynomial that goes through all the points.

%%%%%%%%%%%%%%%%%%%%%%%%%%%%%
\subsection{Recursive Integration}
\label{Section:MultiDimensionalDetails-Recursive}

The integration over smearing factors $c_1$ and $c_3$ is done by a recursive algorithm~\cite{Bradie:943117} which provides a handle on precision.~The algorithm begins by first splitting the integration range into multiple segments and then applying the recursive algorithm to each segment with some pre-specified tolerance level $\varepsilon$.~More explicitly the recursive algorithm proceeds as follows:
\begin{itemize}
\item Apply second order Newton-Cotes quadrature to the whole range to obtain a first order approximation for the integral which we label $I_0$.
\item We then split the segment in half and apply the approximation on each of the half-segments
to obtain a second estimate $I_1 + I_2$.
\item Next we estimate integration error $\delta I$ by comparing $I_0$ and $I_1 + I_2$.
\item If $|\delta I | < \varepsilon$, we terminate the algorithm and use $I_1 + I_2$ as the integral.
\item Otherwise, we repeat the procedure on each of the two half-segments and require that each satisfy a tolerance level of $\varepsilon / 2$.
\end{itemize}
This procedure ensures that the overall error is at most $\varepsilon$, assuming that the error estimation is reasonable, which when the function does not vary too rapidly is a good approximation.~This is aided by the fact that the integral is split into smaller intervals where the function varies less and then a sum is taken over all the intervals.

To assess whether the estimate of the error on the integration is reasonable, we first suppose the step size is small enough that there is not a drastic change of landscape inside each integration segment.~Since the estimator we use is a fourth order method, the difference between the actual integral and the estimator can be written as~\cite{Bradie:943117},
\bea
\int\limits_{segment} f(x) dx = I_0 + k h^4 + O(h^5),
\eea
where $h$ is the size of segment and $k$ is a constant characteristic of the method and
independent of $h$.~Applying the same formula to the total of the two half-segments, we arrive at the following,
\bea
\int\limits_{segment} f(x) dx = I_1 + I_2 + 2 k \left(\dfrac{h}{2}\right)^4 + O(h^5).
\eea
Assuming that the term inside $O(h^5)$ is negligible, by comparing the two formulae
we obtain an estimate of integration error to be,
\bea
\delta I \equiv 2k\left(\dfrac{h}{2}\right)^4 = \dfrac{1}{7} \left(I_1 + I_2 - I_0\right).
\eea
In~\fref{MultiDimensionalDetails-Recursive-Action} we show an example of the integration using the adaptive method on two test functions.~In the left panel we perform integration on a second-order polynomial $f(x) = \frac{1}{75} x^2$, where the method is exact.~In this case the grid spacings are always the same.~In the right panel the integrand is modified to be $f(x) = \frac{1}{75} x^2 + e^{-\frac{1}{2}x^2}$.~We can see that in places where functions are rapidly varying, a greater number of points are used.
\begin{figure}
\centering
\includegraphics[width=0.45\textwidth]{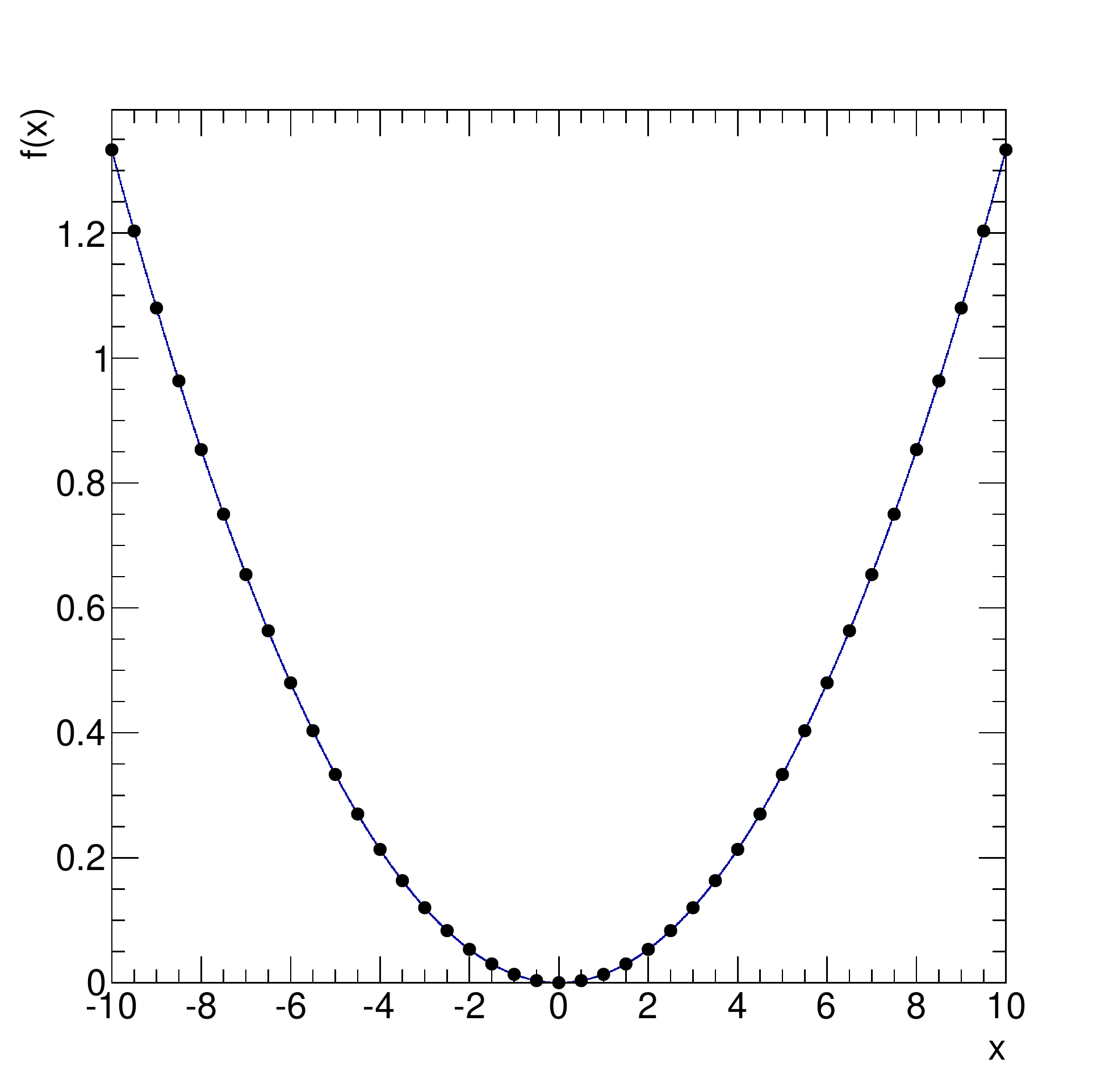}
\includegraphics[width=0.45\textwidth]{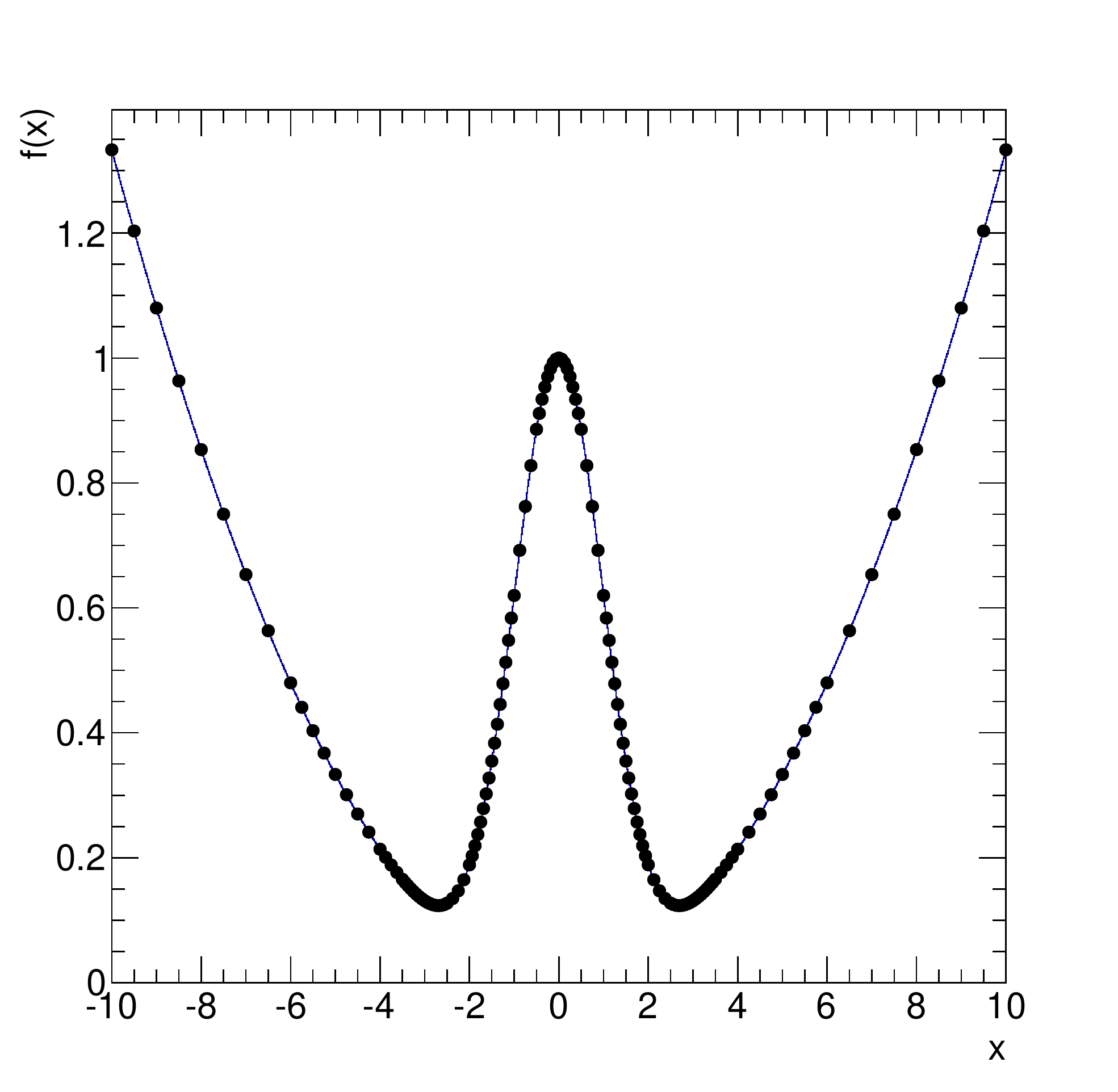}
\caption{Illustration of recursive integration.~On each plot, the line shows true function,
and dots are points evaluated for the integration.~In the left panel we perform integration
on a second-order polynomial $f(x) = \frac{1}{75} x^2$, where the method is exact and therefore,
the grid spacings are always the same.~In the right panel the integrand is modified to be
$f(x) = \frac{1}{75} x^2 + e^{-\frac{1}{2}x^2}$.~We can see that in places where the function
is rapidly varying, more points are used.}
\label{fig:MultiDimensionalDetails-Recursive-Action}
\end{figure}
%

%%%%%%%%%%%%%%%%%%%%%%%%%%%%%%
\subsection{Signal Case Complications}
\label{subsec:MultiDimensionalDetailsSignal}

As discussed in previous sections, in the signal case an anti-correlation between the two
lepton pairs is induced due to the presence of the narrow-width resonance.~We can see the effect on the di-lepton mass due to the narrow-width approximation in~\fref{MultiDimensionalDetails-Signal-Elongation}.~In both plots the inner integral is plotted for the exact same event as a function of $M_1$ and $M_2$.~On the left we do not enforce narrow-width approximation while on the right we require that $\sqrt{\hat{s}} = 125$~GeV.~We see the shape is strongly elongated for the case with narrow-width approximation.
\begin{figure}
\centering
\includegraphics[width=0.45\textwidth]{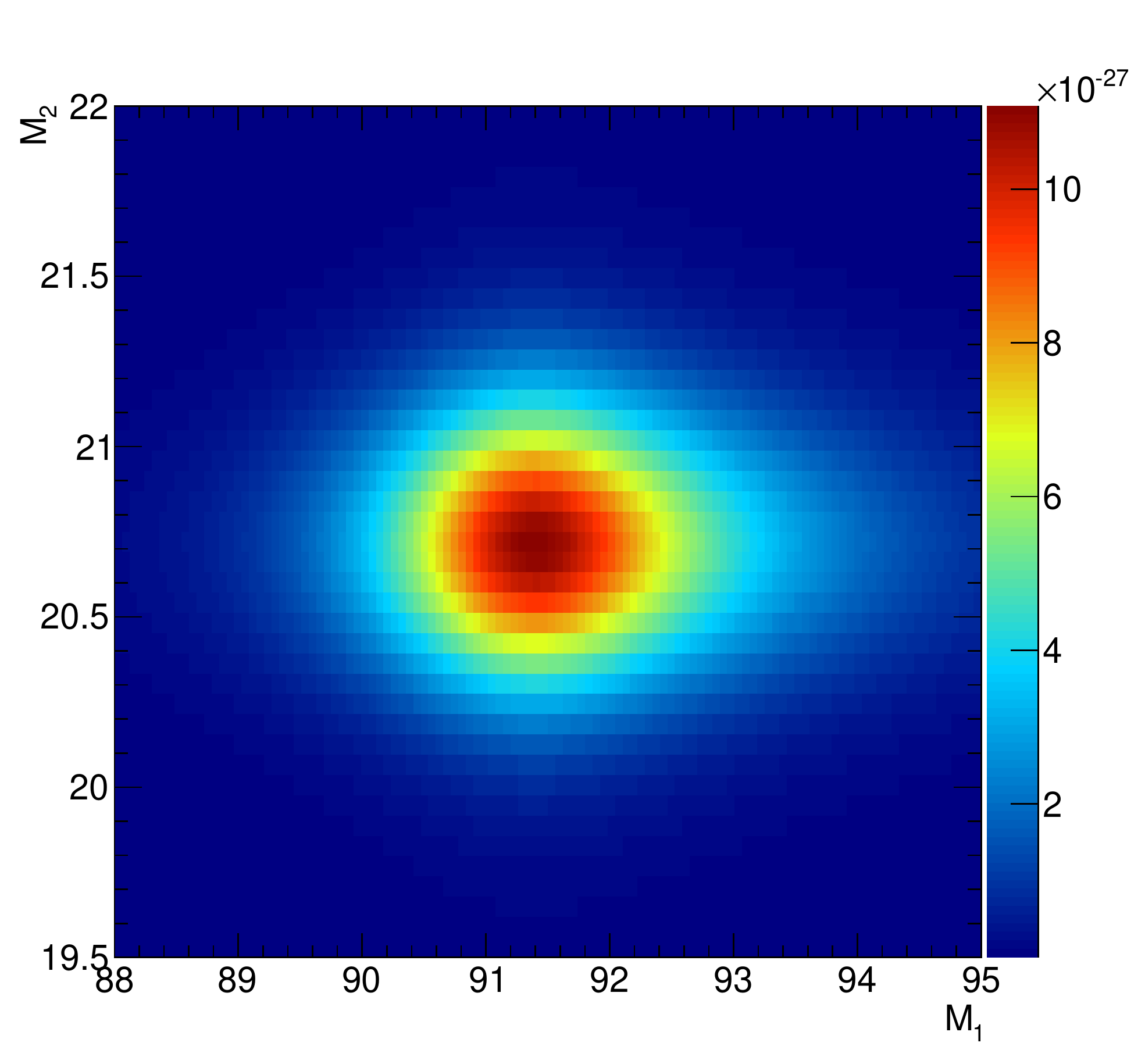}
\includegraphics[width=0.45\textwidth]{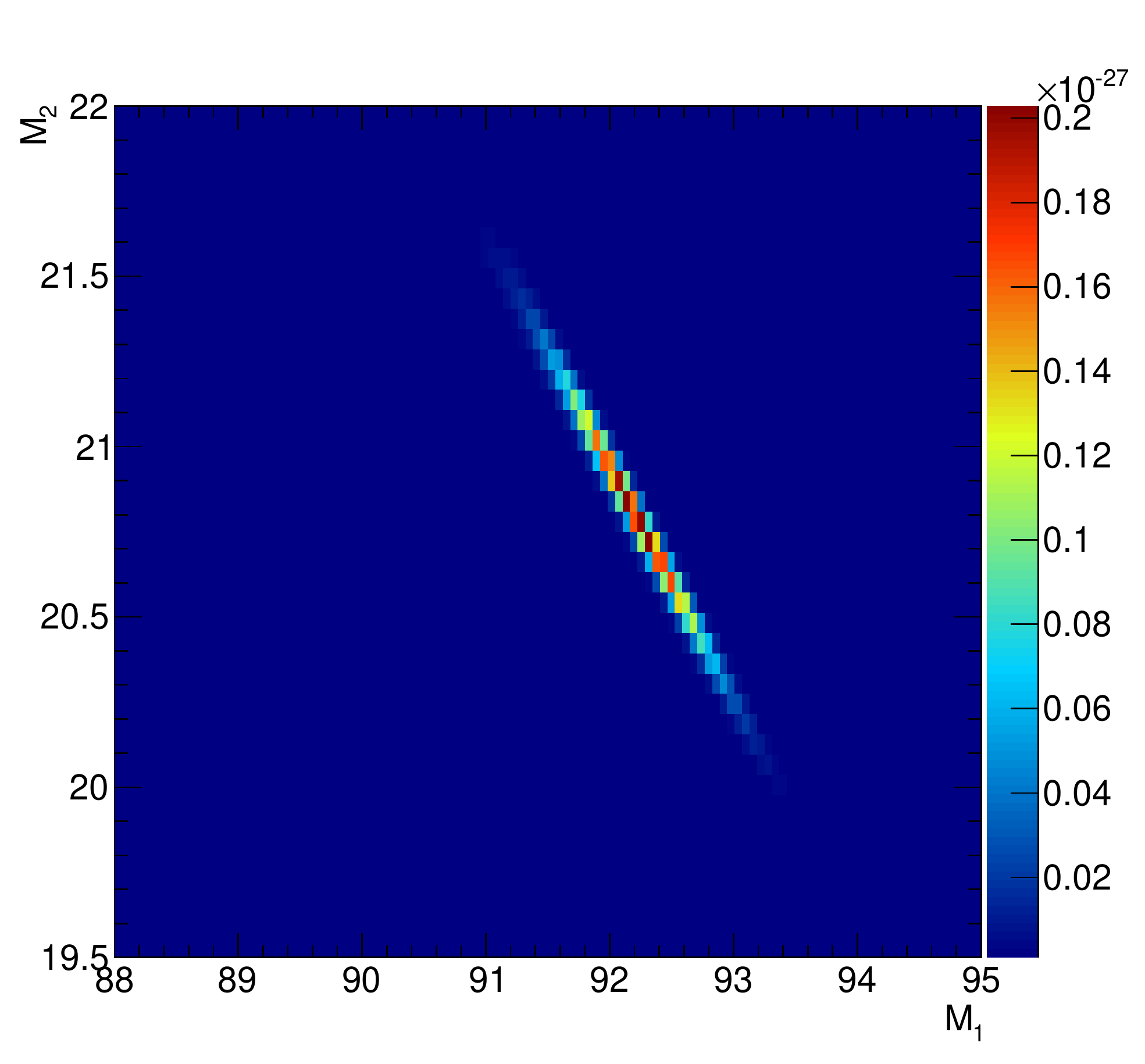}
\caption{Demonstration of the effect on di-lepton masses due to narrow-width approximation.~In both plots the inner integral as a function of $\mlla$ and $\mllb$ is plotted for the exact same event.~On the left we do not enforce narrow-width approximation while on the right we require that $\sqrt{\hat{s}} = 125$~GeV.~We see shape is strongly elongated for the case with narrow-width approximation.}
\label{fig:MultiDimensionalDetails-Signal-Elongation}
\end{figure}

By using a delta function to model width of resonance, there is one less dimension to integrate over as compared to the background case.~While this makes it easier computationally in one respect, an additional complication arises since we have to integrate along a trajectory in which $\mllll$ is kept constant.~Specifically, for a given value of $c_1$ we have to calculate the corresponding value of the other smearing parameter such that $\mlla$, $\mllb$, and $\mllll$ are held constant.~It is easy to keep the di-lepton masses constant since the integral over $c_1$ is the innermost integral and,
\bea
c_1 c_2 \mlla^2 = {m_{12}^R}^2,
\eea
which allows us to simply choose $c_2 = {m_{12}^R}^2 / \mlla^2 c_1$ in order to satisfy the necessary conditions (and similarly for the second lepton pair).~Satisfying the last condition where $\mllll$ is kept constant involves a less trivial calculation.~We begin from the equation,
\bea
\sum_{i>j} c_i^{-1} c_j^{-1} {m_{ij}^R}^2 = \mllll^2 ,
\eea
from which we obtain a quadratic equation for $c_3$ given by,
\bea
\left(R_{34} {m_{14}^R}^2 + c_1^2 R_{34} R_{12} {m_{24}^R}^2\right) c_3^2
- (\mh^2 - \mlla^2 - \mllb^2) c_1 c_3 - \left({m_{13}^R}^2 + c_1^2 R_{12} {m_{23}^R}^2\right) = 0,
\eea
where $R_{12} = \mlla^2 / {m_{12}^R}^2$ and $R_{34} = \mllb^2 / {m_{34}^R}^2$.~One condition where a solution exists for $c_3$ given $c_1$ value is as follows,
\bea
&& \left[\left(\mh^2-\mlla^2-\mllb^2\right)^2
- 4 R_{34} R_{12} {m_{14}^R}^2 {m_{23}^R}^2 
- 4 R_{12} R_{34} {m_{13}^R}^2 {m_{23}^R}^2\right] c_1^2\nonumber\\
&& - 4 R_{34} {m_{14}^R}^2 {m_{13}^R}^2 
- 4 R_{12}^2 R_{34} {m_{24}^R}^2 {m_{23}^R}^2 c_1^4
\geq 0,
\label{eqn:MultiDimensionalDetails-Signal-SolutionExist}
\eea
which is a quadratic function of $c_1^2$.~We first make a few observations based of~\eref{MultiDimensionalDetails-Signal-SolutionExist}.~First of all, the quadratic coefficient is negative definite, indicating that the parabola is curving downwards.~The constant term is also negative
definite.~If there exists a positive solution for $c_1^2$, both solutions are positive.~We also require that the linear coefficient is positive which ensures that if there is a solution, there exists a positive solution.~To summarize, the two conditions that need to be met in order for solutions of the smearing parameters to exist that will meet all mass requirements are given by,
\bea
& &\left(\mh^2-\mlla^2-\mllb^2\right)^2
- 4 R_{34} R_{12} {m_{14}^R}^2 {m_{23}^R}^2 
- 4 R_{12} R_{34} {m_{13}^R}^2 {m_{23}^R}^2
\nonumber\\
&& \phantom{ABCDE} > 8 R_{12} R_{34} 
{m_{13}^R} {m_{14}^R} {m_{23}^R} {m_{24}^R}~~(> 0).
\eea
The first equality is the necessary condition for~\eref{MultiDimensionalDetails-Signal-SolutionExist}
to be true combined with the condition that the linear coefficient is to be positive.~If these conditions are met, all allowed solutions form an ellipsoidal contour in the positive quadrant of the $c_1-c_3$ plane.~Some example ellipses are shown in~\fref{MultiDimensionalDetails-Signal-C1C3Ellipse}.
\begin{figure}
\centering
\includegraphics[width=0.45\textwidth]{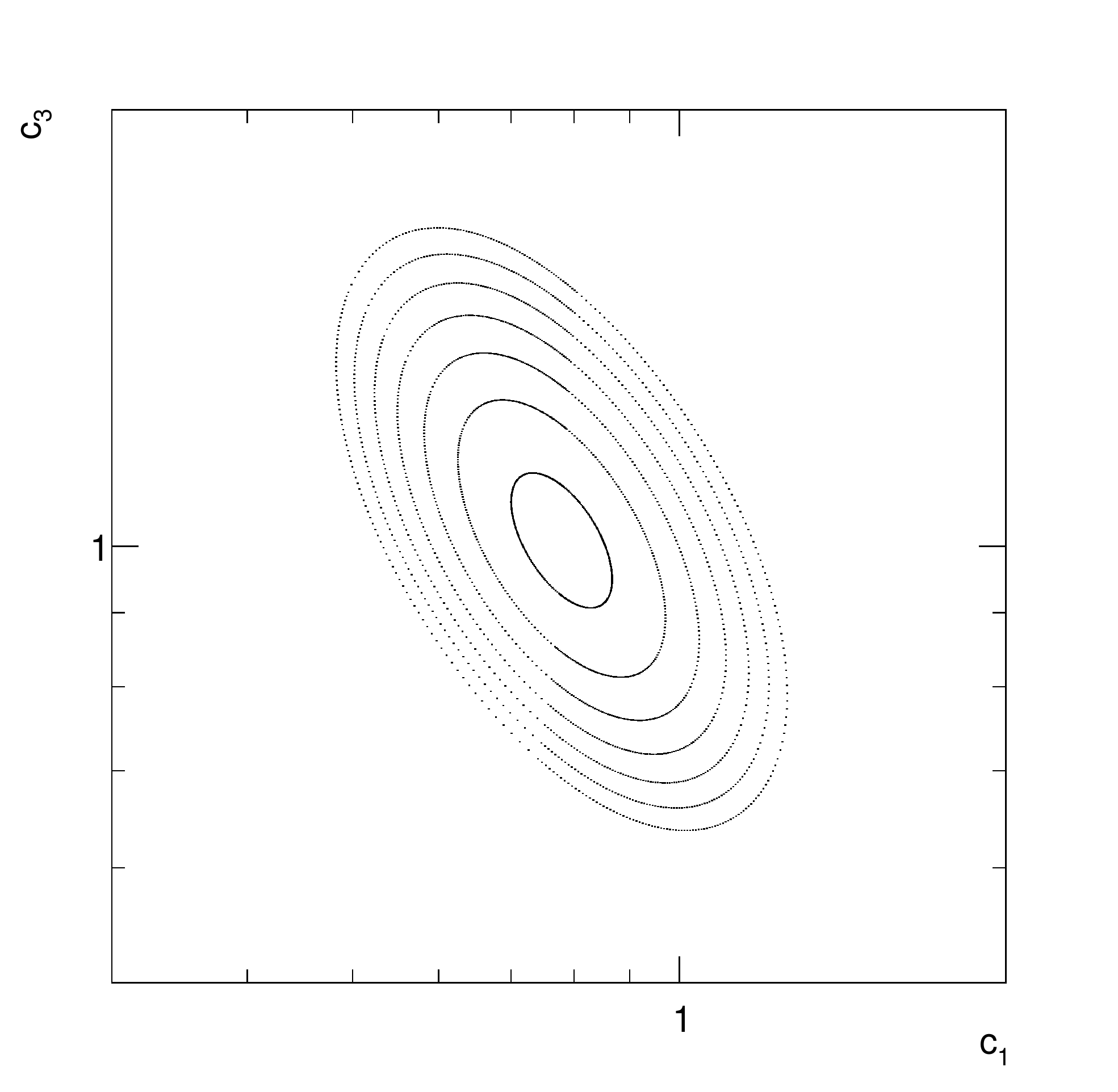}
\includegraphics[width=0.45\textwidth]{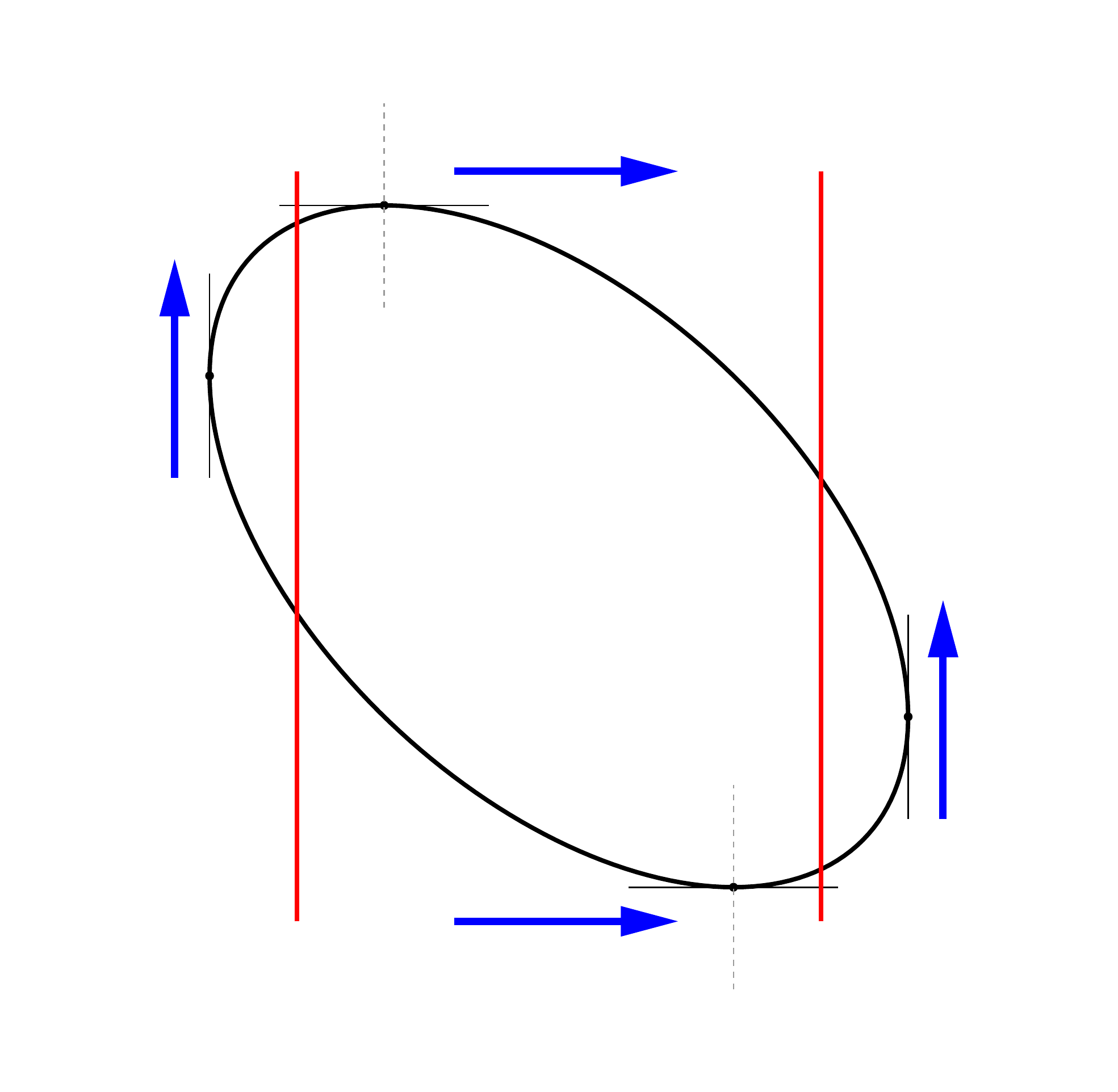}
\caption{Examples of $c_1-c_3$ ellipses for a few mass grid points on an example event are shown in the left panel where each contour represents one mass grid point.~Dots represent possible solutions to all mass constraints.~For each grid point all possible solutions form an ellipsoidal shape, as expected.~In the right we show the slicing of ellipse while doing the integral.~The four extreme points (left, right, up, down) are calculated and the curve is divided into four regions by the red lines going vertically between the extreme points.~In order to avoid artificial infinities during integration we integrate along the horizontal direction for the two segments in the middle and vertically upwards along $c_3$ for the two segments on the side shown by blue arrows.}
\label{fig:MultiDimensionalDetails-Signal-C1C3Ellipse}
\end{figure}

Due to the shape, we need to integrate different regions of the ellipse separately.~To do this we ``slice'' this ellipse into four regions.~First, two points are picked  by calculating the average value of $c_1$ between the leftmost point of the ellipse and the smaller of two points where $dc_3/dc_1 = 0$.~Similarly we can also define another point using the rightmost point and the larger of two points where $dc_3/dc_1 = 0$.~The four regions are thus defined by the two horizontal segments between the two aforementioned points and two vertical segments on the two sides.~An example is shown in~\fref{MultiDimensionalDetails-Signal-C1C3Ellipse}.~We integrate the horizontal segments normally over $c_1$.~For the vertical segments we flip the role of $c_1$ and $c_3$ and integrate along the $c_3$ direction, adjusting $c_1$ so that all mass requirements are met.~This is necessary as the signal Jacobian factor $|{\bf J}^{\vec{M}}_S|$ diverges at the edge of ellipse where the direction of curve approaches vertical.

The complete validation of the full convolution procedure can be found in~\cite{Chen:2014pia,CMS-PAS-HIG-14-014,Thesis}.
%%%%%%%%%%%%%%%%%%%%%%%%%%%%%%%%%%%%%%
\section{Constructing Likelihoods and Maximization}
\label{sec:likelihoods}

To obtain the final likelihoods we must first normalize the detector level differential cross sections to obtain a proper \emph{pdf}.~Once a \emph{pdf} is obtained we can go on to construct the final likelihoods with which multi-dimensional parameter extraction can be perfumed.~In this section we describe how the normalization is performed and sketch how parameter extraction can be done.

%%%%%%%%%%%%%%%%%%%%%%%%%%%%%%%%%%%%%%
\subsection{Normalization}
\label{subsec:MultiDimensionalOverview-Normalization}

As described in the previous section, it is necessary to normalize the \pdf properly in order
to construct the likelihoods.~The normalization can be calculated by integrating the $P(\vec{X}^R|\vec{\mathcal{A}})$
over reconstructed level configurations $\vec{X}^R$:
\bea
N(\vec{\mathcal{A}}) \equiv \int P(\vec{X}^R | \vec{\mathcal{A}}) d\vec{X}^R.
\label{Equation:MultiDimensionalOverview-Normalization-Normalization}
\eea
The evaluation of the convolution integral $P(\vec{X}^R|\vec{\mathcal{A}})$ is relatively computationally intensive and it is not possible to repeat it many different times to get a numerical average of the overall normalization via Monte-Carlo methods.~It is also non-trivial to calculate it via numerical methods as there are 12 dimensions and there is no simple way to cut down the dimensions.~We can, however, evaluate the normalization differently and avoid the numerically intensive part as follows,
\begin{align}
N(\vec{\mathcal{A}}) &= \int P(\vec{X}^R | \vec{\mathcal{A}}) d\vec{X}^R \nonumber\\
 &= \int \left( \int P(\vec{X}^G | \vec{\mathcal{A}}) T(\vec{X}^R | \vec{X}^G) d\vec{X}^G \right) d\vec{X}^R \nonumber\\
 &= \int P(\vec{X}^G | \vec{\mathcal{A}}) \left( \int T(\vec{X}^R | \vec{X}^G) d\vec{X}^R \right) d\vec{X}^G \nonumber\\
 &\equiv \int P(\vec{X}^G | \vec{\mathcal{A}}) \bar{\epsilon}(\vec{X}^G) d\vec{X}^G,
\end{align}
where in the last step we define the ``average efficiency'' given a certain generator level configuration as
$\bar{\epsilon}(\vec{X}^G)$.~It is the average efficiency a certain generator-level event will survive all analysis cuts.
This way we avoid the computationally difficult parts and a straightforward Monte-Carlo algorithm is sufficiently precise.~In~\fref{MultiDimensionalOverview-Normalization-Precision} we show an example of
convergence of the normalization calculation as a function of sample count and CPU time.~We can reach a precision of 0.1\% in a few CPU-hours of run time for all components.
\begin{figure}
\centering
\includegraphics[width=0.6\textwidth]{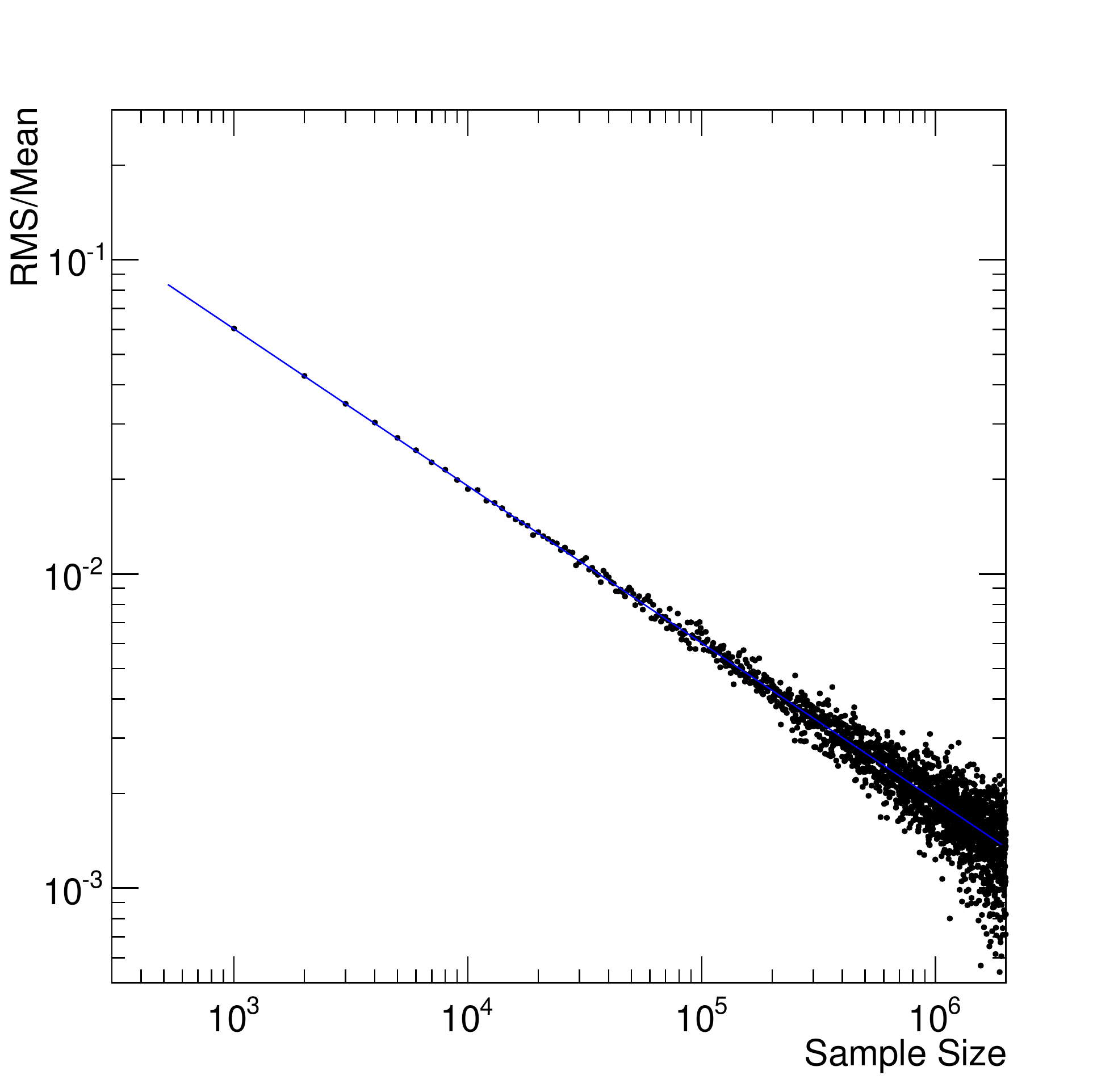}
\caption{Controlling normalization precision}
\label{fig:MultiDimensionalOverview-Normalization-Precision}
\end{figure}
%

%%%%%%%%%%%%%%%%%%%%%
\subsection{Signal Likelihood for Arbitrary Model Point}
\label{subsec:MultiDimensionalOverview-SignalCombination}

Due to the simple (polynomial) dependance of the signal differential cross section on the lagrangian parameters, it is possible to pre-calculate different pieces of the detector level likelihood separately, and later combine them in a trivial way to get the final detector-level likelihood.~The polynomial dependence (quadratic in the simplest case~\cite{Chen:2014pia}) allows us to write the generator-level differential cross section in the form
\bea
P_S(\vec{X}^G | \vec{\mathcal{A}}) = \sum_i f_i(\vec{\mathcal{A}}) P^i_S(\vec{X}^G),
\eea
where the index $i$ runs over different terms in the expression (see~\cite{Chen:2014pia} for description of the different terms) and $f_i(\vec{\mathcal{A}})$ is some polynomial function of the lagrangian parameters.~The convolution integral can then be done on pieces $P^i_S(\vec{X}^G)$ which are independent of $\vec{\mathcal{A}}$ and later combined together:
\begin{align}
P(\vec{X}^R | \vec{\mathcal{A}}) &= 
\int P(\vec{X}^G | \vec{\mathcal{A}}) T(\vec{X}^R | \vec{X}^G) d\vec{X}^G\nonumber\\
&= \int \sum_i f_i(\vec{\mathcal{A}}) P^i_S(\vec{X}^G) T(\vec{X}^R | \vec{X}^G) d\vec{X}^G\nonumber\\
&= \sum_i \left( \int P^i_S(\vec{X}^G) T(\vec{X}^R | \vec{X}^G) d\vec{X}^G \right) f_i(\vec{\mathcal{A}}).
\end{align}
In the case of current analysis where we have calculated up to leading order in these
couplings, the differential cross section is a second-order polynomial.~The same structure applies to the \pdf normalization, which is also a polynomial of parameters we want to measure:
\begin{align}
N(\vec{\mathcal{A}}) &= \int P(\vec{X}^G | \vec{\mathcal{A}}) \bar{\epsilon}(\vec{X}^G) d\vec{X}^G\nonumber\\
&= \sum_i \left( \int P^i_S(\vec{X}^G) \bar{\epsilon}(\vec{X}^G) d\vec{X}^G \right) f_i(\vec{\mathcal{A}}).
\end{align}
Once we have the coefficients pre-calculated, it's straightforward and fast to combine into final likelihood
for arbitrary values of parameters of interest.~We can obtain the background likelihood in a similar manner except the overall normalization does not depend on any undetermined parameters which simples the procedure.~The final signal plus background can then be constructed as described in~\cite{Chen:2014pia,CMS-PAS-HIG-14-014,Thesis}.

%%%%%%%%%%%%%%%%%%%%%%%%%%%%%%%%%%%%%%
\subsection{Maximization of Likelihood and Parameter Extraction}
\label{subsec:maximization}

There are various ways one can go on to perform parameter extraction once the likelihood is constructed.~One can for example simply scan the likelihood for each parameter point as done in~\cite{CMS-PAS-HIG-14-014}.~We instead implement a maximization procedure based on the MINUIT~\cite{James:1994vla} function minimization code which is incorporated into our framework in order to find the maximum of the likelihood.~More specifically, once the likelihood $L(\vec{\mathcal{A}})$ for a particular dataset is obtained, the maximization procedure is utilized in order to obtain the value of the parameters which maximizes the likelihood, which we label $\hat{\mathcal{A}}$.~Thus $\hat{\mathcal{A}}$ represents the most likely value of $\vec{\mathcal{A}}$ for a given dataset which is schematically represented as,
\bea
\label{eqn:max_likelihood}
\frac{\partial L(\vec{\mathcal{A}})}{\partial\vec{\mathcal{A}}} \Big|_{\vec{\mathcal{A}} = \hat{\mathcal{A}}} = 0.
\eea
One important feature of the procedure is that the computationally intensive component of evaluating the likelihood only needs to be done for the events in the final dataset used in the fit for a given experiment.~Therefore the computationally expensive pieces can be calculated on the computing grid prior to the analysis of the data, and the fits for the parameter extraction itself is then completed within a few seconds.~This allows for a great deal of flexibility when fitting the undetermined parameters. 

In practice the maximization in Eq.(\ref{eqn:max_likelihood}) is done by starting from some random initial point in the parameter space and utilizing the built in algorithm in the MINUIT minimization code to efficiently find the maximum.~Of particular importance in this step is ensuring that the point in parameter space that this procedure converges to is actually the \emph{global maximum} and not simply a local maximum, as the statistical fluctuations of a particular dataset can lead to the appearance of multiple local maxima in the likelihood.~This can lead to biases or imprecise estimations of the undetermined parameters.

\begin{figure}
\centering
{\bf~~Finding the Global Maximum}\\
~\\
\includegraphics[width=0.4\textwidth]{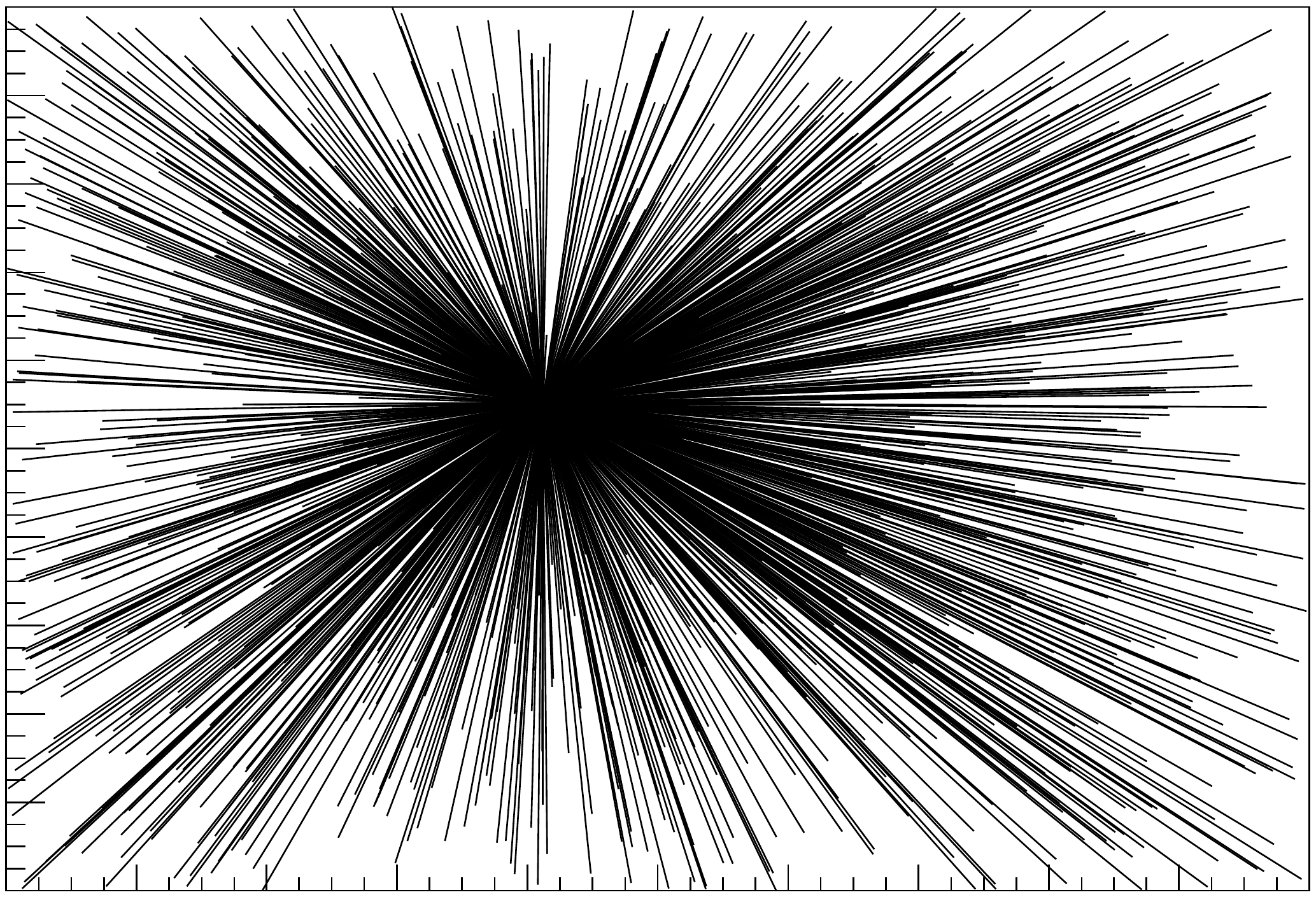}
\includegraphics[width=0.4\textwidth]{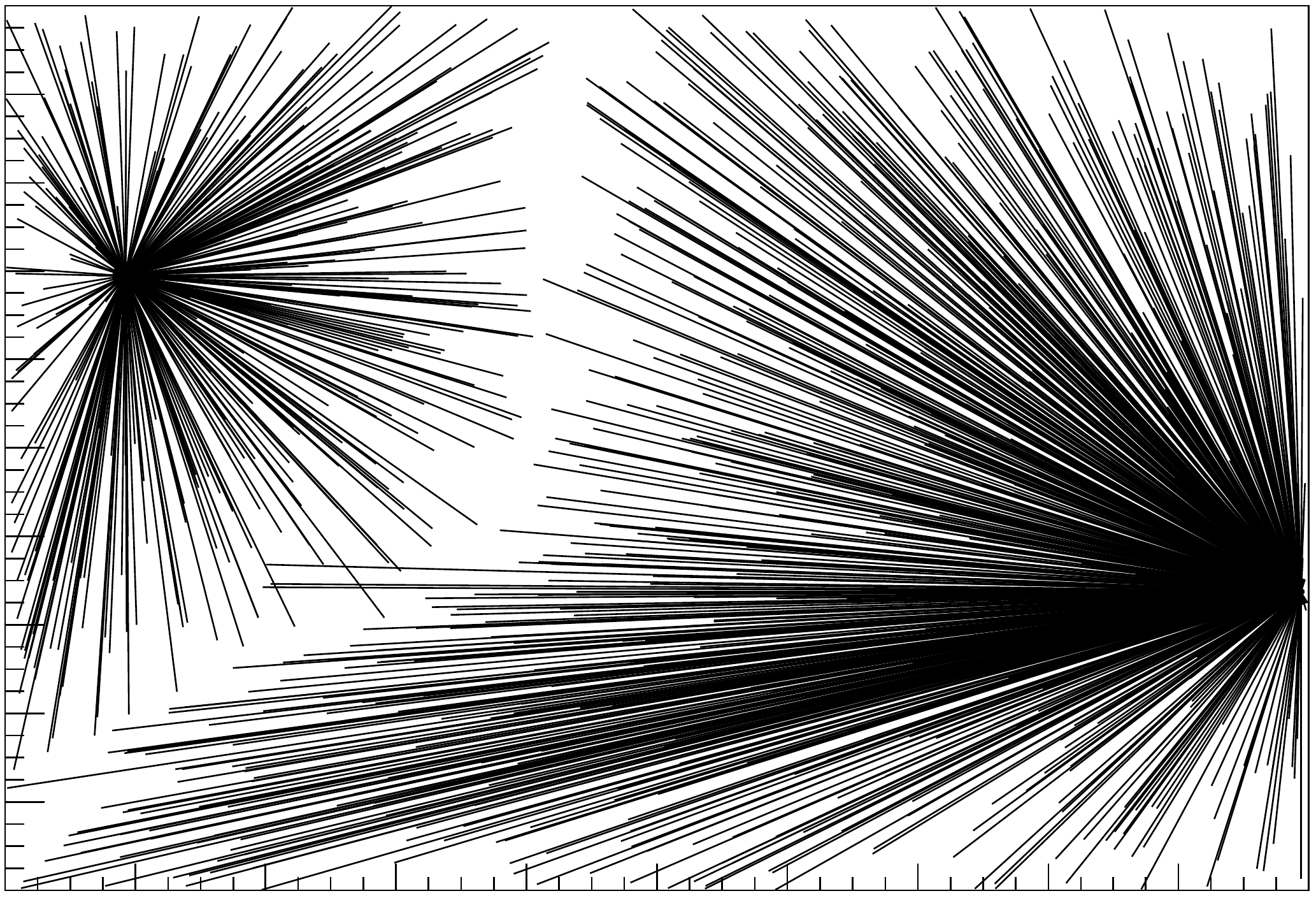}
\caption{`Arrow plots' showing the convergence to the point which maximizes the likelihood starting from a random initial point.~In these the tail of the arrow is at the initial point while the head is at the end point to which the fit converges.~On the left we see that only one end point is found for all initial values.~On the right we see the appearance of two endpoints which depend on the initial value indicating the appearance of multiple maxima.}
\label{fig:localmax}
\end{figure}

We illustrate this effect in Fig.~\ref{fig:localmax} where we show `arrow plots' for an example two-dimensional fit to two different datasets containing the same number of events and same `true' value for the undetermined parameters.~We show a large number of arrows whose tails begin at some initial point in a two dimensional parameter space and whose heads point to the final point reached in the maximization scan.~On the left we see the same endpoint is reached regardless of the initial starting point indicating there is a clear global maximum.~On the right we see two separate accumulations to which the arrow heads point indicating two local maxima.~We have carefully accounted for this effect in our maximization procedure and find a very high convergence rate in general ($\gtrsim 99\%$) to the global maximum of the likelihood.~Various demonstrations of the parameter extraction can be found in~\cite{Chen:2013ejz,Chen:2014gka} at generator level and inin~\cite{CMS-PAS-HIG-14-014,Thesis} for detector level.
%%%%%%%%%%%%%%%%%%%%%%%%%
%%%%%%%%%%%%%%%%%%%%%%%%%

To quantify the uncertainty on the extracted value $\hat{\mathcal{A}}$ we perform a large number of pseudo-experiments $\mathcal{N}$ each containing $N$ events and perform the maximization for each pseudo-experiment.
A distribution for $\hat{\mathcal{A}}$ is obtained with a spread $\sigma$ and average value $\bar{\mathcal{A}}$.~The true value $\mathcal{\mathcal{A}}_o$ will sit within some interval of the extracted value $\hat{\mathcal{A}}$ for a given pseudo experiment and as the number of pseudo experiments is taken to infinity the average value of $\hat{\mathcal{A}}$ will converge to the true value; i.e.~$\bar{\mathcal{\mathcal{A}}} \rightarrow \mathcal{\mathcal{A}}_o$ as $\mathcal{N} \rightarrow \infty$.~Further discussion of the statistical methods used can be found in~\cite{CMS-PAS-HIG-14-014,Thesis}.

%%%%%%%%%%%%%%%%%%%%%%%%%%%%%%%%%%%%%%
\section{Summary}
We have described various technical details of a novel analysis framework introduced in~\cite{Chen:2014pia} to measure properties of the newly-discovered bosonic state at $\sim125$~GeV.~This analysis method allows us to fully utilize the power of the golden channel by constructing a continuous likelihood function in all observables, which is also a continuous function of the parameters of interest.~We have emphasized in particular the details involved in performing the convolution integral which takes one from `truth' level observables to detector level observables.~We have also briefly discussed other aspects of the framework.

This framework is distinctively different compared to other established methods which utilize
templates of discriminants.~These template methods offer many advantages such as simplicity and faster analysis set up time, but are more dependent on availability of Monte-Carlo samples and the choice of discriminant.~Our framework on the other hand, though more complex and requiring (arguably) more computing resources \emph{prior to analyzing data}, offers great speed, flexibility, and a more complete picture on the extracted results when actually performing the parameter extraction on data.~Complete validations of the framework can be found in~\cite{CMS-PAS-HIG-14-014,Thesis}.~With this framework one can then go on to perform a variety of multi-parameter extractions in the $h\to4\ell$ channel with data obtained at the LHC and other future colliders.

%%%%%%%%%%%%%%%%%%%%%%%%%%%%%%%%%%%%%%
\section*{Acknowledgements}
We thank the ATLAS and CMS collaborations for their encouragement and interest in this work.~The work of R.V.M.~is supported by the ERC Advanced Grant Higg@LHC and the Fermilab Graduate Student Fellowship in Theoretical Physics.~Y.C.~is supported by the Weston Havens Foundation and DOE grant No.~DE-FG02-92-ER-40701.~Fermilab is operated by Fermi Research Alliance, LLC, under Contract No.~DE-AC02-07CH11359 with the United States Department of Energy.~This work is also sponsored in part by the DOE grant No.~DE-FG02-91ER40684.~R.V.M is also grateful to the CalTech physics department for their hospitality during which much of this work was done.~This work used the Extreme Science and Engineering Discovery Environment (XSEDE), which is supported by National Science Foundation grant number OCI-1053575.

\bibliographystyle{JHEP}
\bibliography{GoldenChannelBib}

\end{document}